\documentclass[11pt,aps,preprintnumbers,nofootinbib,superscriptaddress,
               notitlepage,longbibliography]{revtex4-2}

\makeatletter

\let\NAT@parse\relax
\expandafter\let\csname ver@natbib.sty\endcsname\relax
\let\textcite\relax

\let\auto@bib\relax
\long\def\addtocontents#1#2{\protected@write\@auxout{\robustify@contents\let\label\@gobble\let\index\@gobble\let\glossary\@gobble}{\string\@writefile{#1}{#2}}}
\makeatother
 
\usepackage[T1]{fontenc}
\usepackage[english]{babel}
\usepackage{amsmath,bm,bbm,physics,dsfont}
\makeatletter
\newtoks\tnsr@acc
\newcount\tnsr@state
\newif\iftnsr@first
\DeclareRobustCommand*{\tensor}[2]{#1\tnsr@firsttrue\tnsr@state\z@\tnsr@acc{}\tnsr@scan#2\tnsr@end}
\def\tnsr@scan{\@ifnextchar^{\tnsr@sup}{\@ifnextchar_{\tnsr@sub}{\tnsr@finish}}}
\def\tnsr@sup^#1{\ifnum\tnsr@state=\tw@\tnsr@flush\fi\tnsr@state\@ne\tnsr@acc\expandafter{\the\tnsr@acc#1}\tnsr@scan}
\def\tnsr@sub_#1{\ifnum\tnsr@state=\@ne\tnsr@flush\fi\tnsr@state\tw@\tnsr@acc\expandafter{\the\tnsr@acc#1}\tnsr@scan}
\def\tnsr@flush{\iftnsr@first\tnsr@firstfalse\else{}\fi
  \ifnum\tnsr@state=\@ne^{\the\tnsr@acc}\else_{\the\tnsr@acc}\fi\tnsr@acc{}}
\def\tnsr@finish\tnsr@end{\ifnum\tnsr@state>\z@\tnsr@flush\fi}
\makeatother
\usepackage{stix}
 
\usepackage{graphicx}
\usepackage{verbatim}
\usepackage{csquotes}
\usepackage{simplewick}
\usepackage[usenames,dvipsnames,svgnames]{xcolor}
\usepackage[normalem]{ulem}
\usepackage{hyperref}
\usepackage{cleveref}
\usepackage{orcidlink}

 
\usepackage[style=numeric-comp,sorting=none,backend=biber]{biblatex}
\defbibheading{bibliography}[\refname]{\section*{#1}}
\usepackage{citemsp}
 
\definecolor{IAN}{RGB}{1,80,158}

\usepackage[framemethod=default]{mdframed}
\newmdenv[skipabove=7pt,
skipbelow=7pt,
rightline=false,
leftline=false,
topline=false,
bottomline=false,
backgroundcolor=gray!10,
linecolor=gray,
innerleftmargin=5pt,
innerrightmargin=5pt,
innertopmargin=5pt,
innerbottommargin=5pt,
leftmargin=0cm,
rightmargin=0cm,
linewidth=4pt]{eBox}
 
\newcommand{\md}{\mathrm{d}}

\begin{document}
 
\title{Hamiltonian analysis of disformally coupled fermions in scalar-tensor theories}

\author{Alexander Ganz\,\orcidlink{0000-0001-7939-9058}}
\affiliation{Institute for Theoretical Physics, Leibniz University Hannover, Appelstra{\ss}e 2, 30167 Hannover, Germany}
\email{alexander.ganz@itp.uni-hannover.de}

\author{Apostolos Tsampodimos\,\orcidlink{0009-0001-0230-5647}}
\affiliation{Center for Physical Sciences and Technology, Sauletekio 3, 10257 Vilnius, Lithuania}
\email{apostolos.tsampodimos@ftmc.lt}

    \begin{abstract}
        We study disformal transformations of $\mathcal{N}$ scalar fields coupled to a Dirac fermion. The resulting axial coupling generically introduces higher derivatives in the scalar field equations of motion; for field-dependent free functions, these are absent if and only if the disformal transformation acts along a single field-space direction. We show that these higher derivatives do not signal an Ostrogradsky ghost. After the transformation, the fermion registers the derivatives of the matter-frame tetrad only through the axial current, so that every coupling to the derivatives of the scalar fields multiplies a fermion bilinear, which is Grassmann even and nilpotent. Such terms do not contribute to the body of the constraint matrix, and at each level of the Grassmann expansion the field equations are linear, with the principal part of the theory without fermions. Through a Hamiltonian analysis, we show that the number of degrees of freedom is that of the bosonic theory plus that of the fermion, both in the Einstein frame, for free functions depending on the scalar fields and their kinetic matrix, and for minimally coupled fermions in multi-field DHOST theories. The same mechanism applies to disformal transformations with higher derivatives of a single scalar field, for which fermions impose no condition beyond those required by bosonic matter.
    \end{abstract}

    \maketitle

    \section{Introduction}
    \label{sec:intro}

        While the range of modifications to General Relativity (GR) is vast (see \cite{Tsujikawa:2010zza} for a review), one of the most popular approaches is scalar-tensor theories. In order to study these models in a unified manner, there has been a focus on constructing general scalar-tensor theories with second-order equations of motion, known as Horndeski theories \cite{Horndeski:1974wa,Deffayet:2011gz,Kobayashi:2019hrl}. 
    
        Since the equations of motion are second-order, such theories are free of Ostrogradsky instabilities~\cite{Ostrogradsky:1850fid}, which are generally present in higher-derivative theories and would otherwise render the theory unstable (see, e.g., Refs.~\cite{Woodard:2015zca,Ganz:2020skf,deRham:2016wji} and references therein). However, it was later realized that the restriction to second-order equations of motion is unnecessarily strong: the absence of Ostrogradsky ghosts requires only that the kinetic matrix is degenerate, rather than that the equations of motion be strictly second-order. Relaxing the restriction accordingly leads to Degenerate Higher Order Scalar-Tensor (DHOST) theories~\cite{Gleyzes:2014dya,Gleyzes:2014qga,Langlois:2015cwa,Langlois:2015skt,Crisostomi:2016czh,BenAchour:2016fzp}.
    
        Disformal transformations \cite{auth:BekensteinOriginalDisformal} have played a crucial role in exploring these models. Using these transformations one can shift higher derivatives in the gravity sector to disformal couplings in the matter sector \cite{Brax:2012hm}. As invertible disformal transformations do not change the number of degrees of freedom \citemsp{Domenech:2015tca,Takahashi:2017zgr/III.B}\footnote{The per-key citation locators used throughout this work are produced with the \texttt{citemsp} package, found in \url{https://github.com/ApostolosTsampodimos/citemsp-Latex-package}.}, one can use disformal transformations to construct viable modified gravity models, assuming the disformal coupling itself does not lead to ghost instabilities. In Refs.~\cite{auth:Takahashi_Invertible_Disformal,Takahashi:2022mew,Takahashi:2023vva,auth:Takahashi_Matter_Coupling,Naruko:2022vuh} the authors constructed ``disformal Horndeski'' theories by including higher order derivatives of the scalar field into the disformal transformation. The coupling to matter, however, can break the degeneracy of the gravitational sector~\cite{auth:Deffayet_GarciaSaenz,Takahashi:2022mew,Naruko:2022vuh}. Fermions seemingly provide a stronger constraint than bosons, as they couple directly to the spin connection and, therefore, to the derivative of the tetrad, and it was argued that their coupling breaks the degeneracy condition and leads to ghosts unless the disformal transformation is further restricted~\citemsp{auth:Takahashi_Matter_Coupling/III.B/e51,auth:Ikeda_Consistency_Couplings}, and consistency conditions for the disformal couplings to fermions were derived in~\citemsp{auth:Domenech_consistency_conditions/IV.A/e82}. More generally, ghost-free derivative interactions between bosons and fermions were studied in~\cite{auth:Kimura_Boson_Fermion,auth:Kimura_Scalar_Fermion}, where the absence of fermionic ghosts was shown to require first-order equations of motion for the fermions~\citemsp{auth:Kimura_Boson_Fermion/IV.C}. Similar concepts of disformal transformation with higher derivatives have also been studied in \cite{Babichev:2021bim,Domenech:2019syf,Domenech:2023ryc,Babichev:2024eoh}.
        
        Another direction for generalizing scalar-tensor theories is multi-field extensions. These are often motivated by higher-dimensional compactifications in string theory and have been extensively studied in the context of inflation \cite{Dimopoulos:2005ac,Kanti:1999vt,Kanti:1999ie,Silverstein:2003hf,Alishahiha:2004eh}. As in the single field case, one can consider multi-field derivative couplings to gravity. For example,~\cite{Padilla:2012dx} proposes a multi-field Galileon model, conjectured to be the most general Lagrangian leading to second-order equations of motion. However, it was later shown in~\cite{Kobayashi:2013ina} that these models constitute only a subset, as they do not include multi-field DBI Galileon models~\cite{Hinterbichler:2010xn,Renaux-Petel:2011rmu} (these are based on the unification of single-field DBI and Galileon models \cite{deRham:2010eu}). For the case of two scalar fields,~\cite{Ohashi:2015fma} derived the most general equations of motion, and subsequently~\cite{Horndeski:2024hee} derived the corresponding Lagrangian, which we refer to as two-field Horndeski. The construction of the most general multi-field Horndeski models is expected to be linked to multi-field disformal transformations~\citemsp{auth:Watanabe/IV}. In \cite{Katayama:2025hnd} the author proposed to define Horndeski theories via multi-field disformal transformations, arguing that the most general scalar-tensor theory with second-order equations of motion is characterized by closure under multi-field disformal transformations with scalar-field dependent free functions, analogous to the single scalar field case. The degeneracy conditions of multi-field higher-order scalar-tensor theories have recently been derived in~\cite{BouzariNezhad:2026zsj}.
        
        In \citemsp{auth:Domenech_consistency_conditions/III.A/e38} it has been shown that two-field disformal couplings to fermions generally lead to higher derivatives in the equations of motion for the two scalar fields, which could hint at Ostrogradsky ghost instabilities.  This appears to contradict the statement that Horndeski theories can be defined via the invariance under disformal transformations. However, a full Hamiltonian analysis to precisely count the number of degrees of freedom was still missing. In this work we provide it and show that there is no contradiction. After an arbitrary disformal transformation, the fermion registers the derivatives of the matter-frame tetrad only through the totally antisymmetric part of the spin connection, which couples to the axial current $J_5^\alpha$. All couplings between the fermion and the derivatives of the scalar fields, and in particular those responsible for the higher derivatives, are therefore proportional to a fermion bilinear, which is Grassmann even but nilpotent. Following~\cite{auth:Deffayet_GarciaSaenz}, we show that such terms modify neither the body of the constraint matrix nor the principal part of the field equations at any level of the Grassmann expansion, so that the theory propagates the degrees of freedom of the bosonic sector plus those of the fermion. The conditions for manifestly second-order equations of motion in the Einstein frame, which we derive for $\mathcal{N}$ fields and which restrict the transformation to a single field-space direction, are thus not required for the absence of ghosts. The same mechanism applies to multi-field DHOST theories and to disformal transformations with higher derivatives, where it removes the additional conditions imposed by fermions in~\cite{auth:Takahashi_Matter_Coupling,auth:Domenech_consistency_conditions}.
        
        The paper is structured as follows. In~\cref{sec:disformal_transformations} we review disformal couplings to fermions. In~\cref{sec:Grassmann} we identify the Grassmann structure of the induced couplings, which we use in~\cref{subsec:DHOST} to count the degrees of freedom for $\mathcal{N}$ fields in three equivalent ways: through a Hamiltonian analysis in the Einstein frame, through the level decomposition of the field equations and through a Hamiltonian analysis of multi-field DHOST theories in the Jordan frame. We discuss disformal transformations with higher derivatives in~\cref{sec:higher_derivative}. Finally, we provide a short discussion of our results in~\cref{sec:concl}. Technical details, including the conditions for manifestly second-order equations of motion in the Einstein frame, are provided in the appendix.

    \section{Disformal transformations}
    \label{sec:disformal_transformations}

            In what follows, we work in $(\mathcal{M},\mathbf{g})$: a smooth, oriented and
            4-dimensional manifold, with the signature $(-,+,+,+)$. We use an \emph{orthonormal
            coframe} $\{\theta^a\}_{a=0,1,2,3} \subset \Omega^1(\mathcal{M})$, and its dual frame
            $\{e_a\}$, such that the metric decomposes pointwise as
                \begin{equation}
                    \mathbf{g} = \tensor{\boldsymbol{\eta}}{_a_b}\theta^a\otimes\theta^b \quad,\quad
                    \theta^a(e_b) = \tensor{\delta}{^a_b}\,.
                \end{equation}
            Throughout, lower-case latin indices $a,b,c,\ldots = 0,1,2,3$ are \emph{Lorentz frame
            indices}, raised and lowered with $\boldsymbol{\eta}$; Greek indices
            $\mu,\nu,\rho,\ldots = 0,1,2,3$ are \emph{spacetime coordinate indices}, raised and lowered
            with $\mathbf{g}$; and capital Latin indices $I,J,K,\ldots = 1,\ldots,\mathcal{N}$ denote the fields.
             
            A \emph{disformal transformation} of the coframe is specified by two data: a
            \emph{conformal factor} $C \in C^\infty(\mathcal{M})$ with $C(p) \neq 0$ for all
            $p \in \mathcal{M}$; and a \emph{disformal map}
            $\mathcal{U} \in \Gamma\bigl(\mathrm{End}(T\mathcal{M})\bigr)$ of the form
                \begin{equation}\label{eq:U-def}
                    \tensor{\mathcal{U}}{^a_b} = \tensor{\delta}{^a_b} + \tensor{\mathbf{L}}{^a_b}\,,
                \end{equation}
            where $\tensor{\mathbf{L}}{_a_b} = \tensor{\boldsymbol{\eta}}{_a_c}\tensor{\mathbf{L}}{^c_b}$
            is a smooth symmetric tensor field, the pure \emph{disformal tensor}. The transformed
            coframe $\tilde{\theta}$ -- the \emph{matter or Jordan frame} -- is
                \begin{equation}\label{eq:tilde_theta}
                    \tilde{\theta}^a = C\,\tensor{\mathcal{U}}{^a_b}\,\theta^b
                    = C\bigl(\tensor{\delta}{^a_b} + \tensor{\mathbf{L}}{^a_b}\bigr)\theta^b\,,
                \end{equation}
            and we call the $\theta$-coframe the \emph{gravity or Einstein frame}. In this work, as
            in~\citemsp{auth:Domenech_consistency_conditions/III/e19}, we consider the particular case of multi-disformal transformations~\citemsp{auth:Watanabe/III/e3.1}
                \begin{equation}\label{eq:L-scalar}
                    \mathbf{L} = \tensor{c}{_I_J}\,\md\phi^I\otimes\md\phi^J
                    \quad\Longrightarrow\quad
                    \tilde{\mathbf{g}} = C^2\,\bigl[\mathbf{g} + \tensor{G}{_I_J}\,\md\phi^I\otimes\md\phi^J\bigr]\,,
                \end{equation}
            where we introduced the field-space kinetic matrix
            $\tensor{X}{^I^J}=\tensor{g}{^\mu^\nu}\tensor{\phi}{^I_\mu}\tensor{\phi}{^J_\nu}$ with $\tensor{\phi}{^I_\mu}=\nabla_\mu\phi^I$, and the
            field-space metric~\citemsp{auth:Domenech_consistency_conditions/III/e26}
                \begin{equation}\label{eq:G-c-relation}
                    \tensor{G}{_I_J} = 2\tensor{c}{_I_J} + \tensor{c}{_I_K}\,\tensor{X}{^K^L}\,\tensor{c}{_L_J}\,.
                \end{equation}
            The functions $C$ and $\tensor{c}{_I_J}$ may depend on $\phi^K$ and $\tensor{X}{^K^L}$. In~\cref{sec:higher_derivative} we further consider disformal transformations of a single scalar field which contain higher derivatives. We assume a Levi-Civita connection both before and after the transformation. On the frame
            bundle a connection is specified by an $\mathfrak{so}(1,3)$-valued 1-form
            $\tensor{\varpi}{^a_b}$; metric compatibility and vanishing torsion hold in both frames.

            For later purposes it is useful to introduce $\mathbf{P} = \tensor{d}{_I_J}\, \md \phi^I \otimes \md \phi^J$ such that 
            \begin{align}
            \label{eq:Inverse_relation}
                \tensor{c}{_I_J} + \tensor{d}{_I_J} + \tensor{c}{_I_K} \tensor{X}{^K^L} \tensor{d}{_L_J} =0~,
            \end{align}
            so that $\tensor{(\mathcal{U}^{-1})}{^\mu_\nu} = \tensor{\delta}{^\mu_\nu} + \tensor{d}{_I_J}\tensor{\phi}{^I^\mu}\tensor{\phi}{^J_\nu}$.

        \subsection{The gravitational and matter actions}
        \label{subsec:action}

            We take the total action to split into a gravitational and a matter sector
                \begin{equation}\label{eq:total-action}
                    S = S_{\mathrm{grav}}[\mathbf{g},\phi^I] + S_{\mathrm{Dirac}}[\tilde{e},\Psi]\,,
                \end{equation}
            in which the disformal fields $\phi^I$ enter the gravitational sector, while matter
            couples to them only through the matter-frame tetrad $\tilde{e}$ of~\eqref{eq:tilde_theta}. We demand the equations of motion of the gravitational sector to be second-order, avoiding any Ostrogradsky ghosts, and diffeomorphism invariance. Concretely, $S_{\mathrm{grav}}$ is a general multi-field Horndeski theory.
             
            We represent matter as a Dirac theory -- in the corresponding matter frame -- in the Hermitian formulation
                \begin{equation}\label{eq:dirac-matter}
                    S_{\mathrm{Dirac}} = -\int\!\md^4x\,\det\tilde{e}\left[ \frac{1}{2}\,\tilde{\bar{\Psi}}\,\tilde{\gamma}^\mu\overset{\leftrightarrow}{\tilde{\nabla}}_\mu\tilde{\Psi} + m\,\tilde{\bar{\Psi}}\tilde{\Psi}\right], \qquad \tilde{\bar{\Psi}}\tilde{\gamma}^\mu\overset{\leftrightarrow}{\tilde{\nabla}}_\mu\tilde{\Psi} \equiv \tilde{\bar{\Psi}}\tilde{\gamma}^\mu\tilde{\nabla}_\mu\tilde{\Psi} - \tilde{\nabla}_\mu\tilde{\bar{\Psi}}\,\tilde{\gamma}^\mu\tilde{\Psi},
                \end{equation}
            where $\tilde{\gamma}^\mu=\tensor{\tilde{e}}{^\mu_a}\gamma^a$ and $\tilde{\nabla}_\mu$ carries the spin connection $\tilde{\varpi}[\tilde{e}]$. Fermions coupling to tetrads, through the spin connection, rather than the metric, allows them to register the local Lorentz frame, which makes their response to a disformal transformation richer than that of bosons~\citemsp{auth:Takahashi_Matter_Coupling/III.B,auth:Deffayet_GarciaSaenz/IV}.
             
            Substituting $\tilde{e} = C\,\mathcal{U}\,e$ and expanding the spin connection, the difference $\mathbf{Z}=\tilde{\varpi}-\varpi$ is algebraic in $\nabla\mathcal{U}$. The fermion couples to $\mathbf{Z}$ \emph{only} through its totally antisymmetric part, because the relevant gamma-matrix structure is~\citemsp{Domenech:2025eii/III.B/e25,auth:Domenech_consistency_conditions/AA/eA9}
                \begin{equation}\label{eq:gamma-identity}
                    \{\gamma^a,\tensor{\Sigma}{^b^c}\} = -\tensor{\epsilon}{^a^b^c_d}\,\gamma^d\gamma^5,
                \end{equation}
            so that the symmetric and trace parts of $\mathbf{Z}$ contract to zero against the Dirac bilinear. More generally, in the Hermitian formulation the spin connection enters only through its totally antisymmetric part, contracted with the axial current, and the torsion-free condition $\md\tilde{\theta}^a = -\tensor{\tilde{\varpi}}{^a_b}\wedge\tilde{\theta}^b$ gives
                \begin{equation}\label{eq:twist}
                    \tensor{\tilde{\varpi}}{_a_b_c}\,\tilde{\theta}^a\wedge\tilde{\theta}^b\wedge\tilde{\theta}^c = \tensor{\eta}{_a_b}\,\tilde{\theta}^a\wedge\md\tilde{\theta}^b\,,
                \end{equation}
            with $\tensor{\tilde{\varpi}}{_a_b}=\tensor{\tilde{\varpi}}{_a_b_c}\,\tilde{\theta}^c$. The fermion therefore registers the derivatives of the matter-frame coframe only through the 3-form~\eqref{eq:twist}, contracted with the axial current. Carrying this out, the action takes the form (see~\citemsp{Domenech:2025eii/III.B/e27,auth:Domenech_consistency_conditions/II.A/e15} for more details)
                \begin{equation}\label{eq:dirac-gravity}
                    S_{\mathrm{Dirac}} = -\int\!\md^4x\,\det e\, C^3 \det \mathcal{U}\left[ \frac{1}{2}\,\left( \tensor{\delta}{^\mu_\nu} + \tensor{d}{_I_J} \tensor{\phi}{^I^\mu} \tensor{\phi}{^J_\nu}  \right)\bar{\Psi}\gamma^\nu\overset{\leftrightarrow}{\nabla}_\mu\Psi + C\,m\,\bar{\Psi}\Psi + i\,M_\alpha J_5^\alpha\right],
                \end{equation}
            with $J_5^\alpha=\bar{\Psi}\gamma^\alpha\gamma^5\Psi$ the axial current, and the axial coupling~\citemsp{Domenech:2025eii/III.A/e22,auth:Domenech_consistency_conditions/II.A/e17}
                \begin{equation}\label{eq:M-def}
                    M^\alpha=\tfrac{1}{4}\,\tensor{\epsilon}{^\alpha^\beta^\gamma^\delta} \tensor{(\mathcal{U}^{-1})}{^\mu_\delta}\tensor{(\mathcal{U}^{-1})}{^\nu_\gamma}\nabla_\mu\tensor{\mathcal{U}}{_\beta_\nu}~.
                \end{equation}
            Rescaling the spinor by the effective conformal factor
                \begin{equation}\label{eq:spinor-rescale}
                    \Psi = C^{3/2}\sqrt{\det\mathcal{U}}\;\tilde{\Psi}\,,
                \end{equation}
            one can absorb the common prefactor as seen in~\citemsp{Domenech:2025eii/III.B/e26,auth:Domenech_consistency_conditions/II.A/e16}. Since the rescaling depends only on $\phi^I$ and $\md\phi^I$, it does not affect the counting of degrees of freedom. We use it in~\cref{sec:Einstein_frame}, while in~\cref{sec:2} a different rescaling is more convenient.
            
            A local Lorentz rotation on the spinor
                \begin{equation}\label{eq:lorentz-redef}
                    \Psi\to\exp\!\big(\tfrac{i}{2}\tensor{T}{_a_b}\tensor{\Sigma}{^a^b}\big)\Psi,
                \end{equation}
            removes the antisymmetric and trace parts of a general $\mathbf{L}$~\citemsp{Domenech:2025eii/B.2/eB8,auth:Domenech_consistency_conditions/B/eB1}, and since $\mathbf{GL}$ does not have a spinor representation, we cannot remove the coupling to the totally antisymmetric part of $\mathbf{Z}$, proportional to $M_\alpha$. Therefore, these transformations exhaust our options, and $M_\alpha$ encodes the
            symmetric-traceless \emph{stretch} of the frame, equivalently the vorticity carried by
            $\tilde{\mathbf{g}}$. In short, after the allowed redefinitions the effect of a disformal transformation on fermionic matter reduces to the axial coupling.
            Because $M_\alpha$ depends on $\nabla\mathcal{U}$, and hence on second derivatives of the $\phi^I$, it introduces third derivatives into the scalar field equations, discussed in detail in~\citemsp{auth:Domenech_consistency_conditions/III} for two scalar fields. In~\cref{sec:2} we show that, for $\tensor{c}{_I_J}=\tensor{c}{_I_J}(\phi^K)$, they are absent if and only if $\mathrm{rank}(\tensor{d}{_I_J})\le1$, i.e.\ if the transformation acts along a single field-space direction $A^1=n_J\,\md\phi^J$, and that for $\tensor{X}{^K^L}$-dependent transformations along $A^1$ the derivatives of $n_J$ with respect to $\tensor{X}{^K^L}$ must in addition be parallel, as is the case for $n_J=n_J(\phi^K)$. In the following we show that, irrespective of these conditions, they do not revive the Ostrogradsky ghost, since $M_\alpha$ only appears contracted with $J_5^\alpha$.

    \section{Body and soul}
    \label{sec:Grassmann}

        Classical fermions are described by Grassmann-odd fields~\cite{auth:DeWitt_Supermanifolds,auth:Henneaux_Teitelboim}. Denoting by $\zeta^r$, $r=1,2,\ldots$, the generators of a Grassmann algebra, $\zeta^r\zeta^s=-\zeta^s\zeta^r$, every supernumber admits the expansion~\citemsp{auth:Deffayet_GarciaSaenz/IV.A/e67--69}
            \begin{equation}\label{eq:level}
                z = z^{(0)} + z_r\,\zeta^r + \tensor{z}{_r_s}\,\zeta^r\zeta^s + \ldots \equiv \sum_{n\ge0} z^{(n)}\,,
            \end{equation}
        where the level-$n$ component $z^{(n)}$ contains $n$ generators. The level-zero component $z^{(0)}$ is the \emph{body}, and $z_S\equiv z-z^{(0)}$ the nilpotent \emph{soul}. Bosonic variables, such as the tetrad and the $\phi^I$, are Grassmann even and have components at even levels only, while $\Psi$ and $\bar{\Psi}$ are Grassmann odd and have components at odd levels only. In particular, their body vanishes.

        The action~\eqref{eq:dirac-gravity} is bilinear in the spinor. Every coupling between the fermion and the bosonic fields therefore multiplies one of the bilinears $\bar{\Psi}\gamma^\nu\overset{\leftrightarrow}{\nabla}_\mu\Psi$, $\bar{\Psi}\Psi$ and $J_5^\alpha$, which are Grassmann even and, being products of two odd quantities, have vanishing body. The second derivatives of the $\phi^I$ enter only through $M_\alpha J_5^\alpha$, as a consequence of~\eqref{eq:twist}. Hence, every term in the bosonic equations of motion that originates from the fermion, including the third-order derivatives of~\cref{sec:2}, is a soul. In the Hamiltonian language, the fermionic contributions to the momenta and to the Poisson brackets among bosonic constraints are souls, while the brackets between bosonic and fermionic constraints are Grassmann odd and, hence, souls as well.

        The relevance of this observation rests on the following property~\citemsp{auth:DeWitt_Supermanifolds,auth:Henneaux_Teitelboim,auth:Kimura_Boson_Fermion/II.A/e20}: a square matrix $\mathsf{C}=\mathsf{C}^{(0)}+\mathsf{C}_S$ with supernumber entries is invertible if and only if its body $\mathsf{C}^{(0)}$ is. More generally, if $\mathsf{C}^{(0)}$ has constant rank $r$, there exist invertible matrices $\mathsf{L}$ and $\mathsf{R}$ such that
            \begin{equation}\label{eq:normal-form}
                \mathsf{L}\,\mathsf{C}\,\mathsf{R} = \begin{pmatrix}\mathds{1}_r & 0\\ 0 & \mathcal{S}\end{pmatrix}, \qquad \mathcal{S}^{(0)}=0\,.
            \end{equation}
        Indeed, choosing $\mathsf{L}_0$ and $\mathsf{R}_0$ such that $\mathsf{L}_0\mathsf{C}^{(0)}\mathsf{R}_0=\mathrm{diag}(\mathds{1}_r,0)$, the upper-left block of $\mathsf{L}_0\mathsf{C}\mathsf{R}_0$ is $\mathds{1}_r$ plus a soul and thus invertible, and block elimination leaves the Schur complement $\mathcal{S}$, whose entries are sums of products of souls. Consequently, a linear system $\mathsf{C}\,u+b=0$ determines exactly $r$ combinations of the unknowns, while the remaining equations, $\mathcal{S}\,\tilde{u} + \tilde{b} = 0$, reduce to $\tilde{b}^{(0)}=0$ at level zero and do not determine $\tilde{u}^{(0)}$. In the Dirac--Bergmann algorithm, $\mathsf{C}$ is the matrix of Poisson brackets among the primary constraints and $u$ are the Lagrange multipliers, so that the rank of the body fixes the number of multipliers that are determined and of secondary constraints that are generated. Since the fermion only contributes souls to the bosonic block and Grassmann-odd entries to the mixed blocks, $\mathsf{C}^{(0)}$ is block diagonal, with the bosonic block of the theory without fermions and the fermionic block, which depends on the bosonic fields only through the matter-frame tetrad. We apply this in the Einstein frame in~\cref{sec:Einstein_frame} and in the Jordan frame in~\cref{subsec:Hamiltonian,sec:higher_derivative}, while in~\cref{sec:EOM} we give the equivalent argument at the level of the equations of motion.

    \section{Multi-field DHOST}
    \label{subsec:DHOST}

        As in the single-scalar field case, disformal transformations with $\tensor{X}{^K^L}$-dependent free functions can be used to construct multi-field DHOST theories. It is convenient to write the transformation in the inverse direction,
            \begin{align}\label{eq:scalar_field_disformal}
                \mathbf{g} = \tilde{C}^2 \bigl[\tilde{\mathbf{g}} + \tensor{\tilde{G}}{_I_J}\,\md\phi^I\otimes\md\phi^J\bigr]\,, 
            \end{align}
        where $\tensor{\tilde{G}}{_I_J}= \tensor{\tilde{G}}{_I_J}(\phi^K,\tensor{\tilde{X}}{^K^L})$ and $\tilde{C}= \tilde{C}(\phi^K,\tensor{\tilde{X}}{^K^L})$. Inserting~\eqref{eq:scalar_field_disformal} into $S_{\mathrm{grav}}$ defines a theory for $\tilde{\mathbf{g}}$, to which the fermion is minimally coupled. In the absence of fermions, the Einstein frame trivially has second-order equations of motion, as the disformal coupling to standard minimally coupled bosonic matter does not lead to higher derivatives. Therefore, provided the transformation is invertible, which is not guaranteed in general, the theory in the Jordan frame is a multi-field DHOST theory fulfilling the constraint structure discussed in~\citemsp{BouzariNezhad:2026zsj/6}. For $\tensor{\tilde{G}}{_I_J}=\tilde{G}\,\tilde{n}_I\tilde{n}_J$ with $\tilde{n}_J=\tilde{n}_J(\phi^K)$, the conditions of~\cref{sec:2} hold, so that the Einstein frame contains no higher derivatives whereas the Jordan frame does. For generic $\tensor{\tilde{G}}{_I_J}$, higher derivatives appear in both frames, contrary to the single scalar field case. The remaining question is whether the fermion, disformally coupled in the Einstein frame and minimally coupled in the Jordan frame, introduces additional degrees of freedom. We answer it in three equivalent ways: through a Hamiltonian analysis in the Einstein frame in~\cref{sec:Einstein_frame}, through the level decomposition of the field equations in~\cref{sec:EOM}, and through a Hamiltonian analysis of multi-field DHOST theories in the Jordan frame in~\cref{subsec:Hamiltonian}. The first does not require the transformation to be invertible, while the last two apply to any multi-field DHOST theory, whether or not it is disformally related to multi-field Horndeski.

        \subsection{Hamiltonian analysis in the Einstein frame}
        \label{sec:Einstein_frame}

        We first make the argument of~\cref{sec:Grassmann} explicit for the action~\eqref{eq:total-action}. For concreteness we take the gravitational sector to be Einstein gravity with $\mathcal{N}$ scalar fields, with field-space metric $\tensor{h}{_I_J}(\phi^K)$ and potential $V(\phi^K)$; since the argument only uses that the fermion contributes souls, it applies to any multi-field Horndeski theory. Following~\citemsp{auth:Deffayet_GarciaSaenz/II/e3}, we remove the second derivatives by introducing the auxiliary fields $\tensor{A}{^I_\mu}$ and the Lagrange multipliers $\tensor{\lambda}{^\mu_I}$. After the rescaling~\eqref{eq:spinor-rescale}, the action reads
            \begin{equation}\label{eq:S_aux}
                \begin{aligned}
                    S = \int\!\md^4x\,\sqrt{-g}\,\bigg[&\frac{M_{\mathrm{Pl}}^2}{2}R - \frac{1}{2} \tensor{h}{_I_J}\,\tensor{g}{^\mu^\nu}\tensor{A}{^I_\mu} \tensor{A}{^J_\nu} - V + \tensor{\lambda}{^\mu_I}\left(\tensor{\phi}{^I_\mu} - \tensor{A}{^I_\mu}\right)\\
                    &- \frac{1}{2}\tensor{\left(\mathcal{U}^{-1}\right)}{^\mu_\nu}\bar{\Psi}\gamma^\nu\overset{\leftrightarrow}{\nabla}_\mu\Psi - C\,m\,\bar{\Psi}\Psi - i\,M_\alpha J_5^\alpha\bigg],
                \end{aligned}
            \end{equation}
        where $\tensor{\phi}{^I_\mu}$ is replaced by $\tensor{A}{^I_\mu}$ in $C$, $\mathcal{U}$ and $M_\alpha$. We use the ADM decomposition of the tetrad,
            \begin{equation}\label{eq:tetrad_ADM}
                \theta^a = N\,n^a\,\md t + \tensor{e}{^a_i}\left(\md x^i+N^i\,\md t\right),
            \end{equation}
        with lapse $N$, shift $N^i$ and spatial components $\tensor{e}{^a_i}$, where $n^a$ is fixed by $n_a\tensor{e}{^a_i}=0$ and $n_an^a=-1$. The fundamental variables are $N$, $N^i$ and $\tensor{e}{^a_i}$, while the spatial metric $\tensor{\gamma}{_i_j}=\tensor{\eta}{_a_b}\tensor{e}{^a_i}\tensor{e}{^b_j}$ is a composite. Moreover, $n^\mu$ is the future-oriented unit normal, with $n^a=\tensor{e}{^a_\mu}n^\mu$, $\tensor{\gamma}{^\mu_\nu}=\tensor{\delta}{^\mu_\nu}+n^\mu n_\nu$ the projector, $\tensor{K}{_i_j} = \frac{1}{2N}(\tensor{\dot{\gamma}}{_i_j} -D_iN_j - D_jN_i)$ the extrinsic curvature and $D_i$ the connection of $\tensor{\gamma}{_i_j}$. Since $M_\alpha$ is linear in $\tensor{\phi}{^I_\mu_\nu}$, also for $\tensor{X}{^K^L}$-dependent $C$ and $\tensor{c}{_I_J}$, we perform the replacement after the 3+1 decomposition and use $\tensor{\phi}{^I_\mu_\nu}=\tensor{\phi}{^I_\nu_\mu}$, so that, with $\tensor{A}{^I_\star}\equiv n^\mu \tensor{A}{^I_\mu}$ and $\tensor{A}{^I_i}$ the spatial components~\citemsp{auth:Deffayet_GarciaSaenz/II/e8},
            \begin{equation}\label{eq:3+1}
                \begin{aligned}
                    &n^\mu n^\nu\tensor{\phi}{^I_\mu_\nu} \to \tensor{V}{^I_\star} \equiv \frac{1}{N}\left(\tensor{\dot{A}}{^I_\star} - N^iD_i\tensor{A}{^I_\star} - \tensor{A}{^I^i}D_iN\right),\\
                    &n^\mu \tensor{\gamma}{^\nu_i}\tensor{\phi}{^I_\mu_\nu} \to D_i\tensor{A}{^I_\star} - \tensor{K}{_i^j}\tensor{A}{^I_j}\,, \qquad
                    \tensor{\gamma}{^\mu_i}\tensor{\gamma}{^\nu_j}\tensor{\phi}{^I_\mu_\nu}\to D_i\tensor{A}{^I_j} - \tensor{K}{_i_j}\tensor{A}{^I_\star}\,.
                \end{aligned}
            \end{equation}
        In this form no time derivative of $\tensor{A}{^I_i}$ appears. Since the Dirac action is linear in the derivatives of the spinor and of the tetrad, and $M_\alpha J_5^\alpha$ is linear in $\tensor{V}{^I_\star}$ and $\tensor{K}{_i_j}$, the fermionic part of the Lagrangian density takes the form
            \begin{equation}\label{eq:LF}
                \mathcal{L}_F = -\frac{1}{2}\sqrt{\gamma}\left(\bar{\Psi}\mathcal{G}\dot{\Psi} - \dot{\bar{\Psi}}\mathcal{G}\Psi\right) + \sqrt{\gamma}\,\mathcal{J}_I\,\tensor{\dot{A}}{^I_\star} + \sqrt{\gamma}\,\tensor{\mathcal{J}}{_a^i}\,\tensor{\dot{e}}{^a_i} - \mathcal{V}_F\,, \qquad \mathcal{G}\equiv -n_\mu\tensor{\left(\mathcal{U}^{-1}\right)}{^\mu_\nu}\gamma^\nu\,,
            \end{equation}
        where $\mathcal{J}_I$, $\tensor{\mathcal{J}}{_a^i}$ and $\mathcal{V}_F$ are bilinear in the spinor and independent of the velocities. In particular, $\mathcal{J}_I$ arises solely from $M_\alpha J_5^\alpha$ and is proportional to the axial current. The Einstein--Hilbert term takes the standard ADM form,
            \begin{equation}\label{eq:EH_ADM}
                S_{\mathrm{EH}} = \frac{M_{\mathrm{Pl}}^2}{2}\int\!\md^4x\,N\sqrt{\gamma}\left({}^{(3)}\!R + \tensor{K}{_i_j}\tensor{K}{^i^j} - K^2\right),
            \end{equation}
        up to boundary terms, and depends on the velocities of the tetrad only through $\tensor{\dot{\gamma}}{_i_j}=2\tensor{\eta}{_a_b}\tensor{e}{^a_{(i}}\tensor{\dot{e}}{^b_{j)}}$.

        The canonical momenta yield the primary constraints
            \begin{equation}\label{eq:primary_E}
                \begin{aligned}
                    &\pi\approx0\,,\qquad \pi_i\approx0\,,\qquad \tensor{\Lambda}{_a_b}\approx0\,,\qquad \tensor{p}{^i_I}\approx0\,,\qquad \tensor{\chi}{^I_i}\equiv \tensor{A}{^I_i}-D_i\phi^I\approx0\,,\qquad \tensor{C}{^\star_I}\equiv \tensor{p}{_\star_I}-\sqrt{\gamma}\,\mathcal{J}_I\approx0\,,\\
                    &\Theta_\Psi\equiv\pi_\Psi-\frac{1}{2}\sqrt{\gamma}\,\bar{\Psi}\mathcal{G}\approx0\,,\qquad \Theta_{\bar{\Psi}}\equiv\pi_{\bar{\Psi}}-\frac{1}{2}\sqrt{\gamma}\,\mathcal{G}\Psi\approx0\,,
                \end{aligned}
            \end{equation}
        where $\pi$, $\pi_i$, $\tensor{\pi}{_a^i}$, $\tensor{p}{_\star_I}$, $\tensor{p}{^i_I}$, $\pi_\Psi$ and $\pi_{\bar{\Psi}}$ are the momenta conjugate to $N$, $N^i$, $\tensor{e}{^a_i}$, $\tensor{A}{^I_\star}$, $\tensor{A}{^I_i}$, $\Psi$ and $\bar{\Psi}$, the latter two defined with left derivatives and the bracket conventions of~\cite{auth:Henneaux_Teitelboim}, and $\tensor{\Lambda}{_a_b}$ is the generator of local Lorentz transformations, which contains the antisymmetric part of $\tensor{\pi}{_a^i}\tensor{e}{_b_i}$ and the spin density. As in~\citemsp{auth:Deffayet_GarciaSaenz/II.A/n4}, the momentum conjugate to $\phi^I$, $\tensor{p}{_\phi_I}=-\sqrt{\gamma}\,\tensor{\lambda}{^\star_I}$ with $\tensor{\lambda}{^\star_I}\equiv n_\mu\tensor{\lambda}{^\mu_I}$, is used to eliminate $\tensor{\lambda}{^\star_I}$, while $\tensor{\lambda}{^i_I}$ enforces $\tensor{\chi}{^I_i}\approx0$. The symmetric part $\tensor{\pi}{^i^j}\equiv\frac{1}{2} \tensor{e}{^a^{(i}}\tensor{\pi}{_a^{j)}}$ equals $\frac{M_{\mathrm{Pl}}^2}{2}\sqrt{\gamma}\,(\tensor{K}{^i^j}-K\tensor{\gamma}{^i^j})$ plus a soul and is solved for $\tensor{K}{_i_j}$, since the body of the coefficient is invertible. With $\tensor{A}{^I_\star}$ and $E^0\equiv-n_\mu\gamma^\mu$ as variables~\citemsp{auth:Deffayet_GarciaSaenz/IV.A/e70}, the canonical Hamiltonian is, up to primary constraints, $H_C=\int\md^3x\,(N\mathcal{H}_0+N^i\mathcal{H}_i)$, and the secondary constraints $\mathcal{H}_\mu\approx0$ generated by $\pi\approx0$ and $\pi_i\approx0$ are, together with $\pi$, $\pi_i$ and $\tensor{\Lambda}{_a_b}$, first class by diffeomorphism and local Lorentz invariance~\citemsp{auth:Deffayet_GarciaSaenz/IV.A}.

        The matrix of Poisson brackets of the remaining primary constraints, ordered as $(\tensor{C}{^\star_I}$, $\tensor{p}{^i_I}$, $\tensor{\chi}{^I_i}$, $\Theta_\Psi$, $\Theta_{\bar{\Psi}})$, has the body
            \begin{equation}\label{eq:body_E}
                \mathsf{C}^{(0)} = \begin{pmatrix} 0 & 0 & 0\\ 0 & \mathsf{P} & 0 \\ 0 & 0 & \mathsf{F}\end{pmatrix}, \qquad
                \mathsf{P} = \begin{pmatrix}0 & -\mathds{1}_{3\mathcal{N}}\\ \mathds{1}_{3\mathcal{N}} & 0\end{pmatrix}, \qquad
                \mathsf{F} = -\sqrt{\gamma}\begin{pmatrix}0 & \mathcal{G}^{\mathrm{T}}\\ \mathcal{G} & 0\end{pmatrix},
            \end{equation}
        where we used $\{\Theta_\Psi,\Theta_{\bar{\Psi}}\}=-\sqrt{\gamma}\,\mathcal{G}^{\mathrm{T}}$ and $\{\Theta_\Psi,\Theta_\Psi\}=\{\Theta_{\bar{\Psi}},\Theta_{\bar{\Psi}}\}=0$. The other entries are non-vanishing in general: $\{\tensor{C}{^\star_I},\tensor{C}{^\star_J}\}$ and $\{\tensor{C}{^\star_I},\tensor{p}{^i_J}\}$ are bilinear in the spinor, since $\mathcal{J}_I$ depends on $\tensor{A}{^J_\star}$ and $\tensor{A}{^J_i}$, while $\{\tensor{C}{^\star_I},\Theta_\Psi\}$ and $\{\tensor{p}{^i_I},\Theta_\Psi\}$, and similarly for $\Theta_{\bar{\Psi}}$, are linear in the spinor, since $\mathcal{G}$ depends on $\tensor{A}{^J_\star}$ and $\tensor{A}{^J_i}$. None of them contributes to the body. Since $\mathcal{G}^2 = C^2\,\tensor{\tilde{g}}{^\mu^\nu}n_\mu n_\nu\,\mathds{1}$, $\mathsf{F}$ is invertible provided the constant-time hypersurfaces are spacelike also with respect to $\tilde{\mathbf{g}}$, which we assume. Since the action is linear in the velocities of the spinor, which are not multiplied by bosonic velocities, $\Theta_\Psi$ and $\Theta_{\bar{\Psi}}$ provide the eight real primary constraints required to avoid fermionic ghosts~\citemsp{auth:Kimura_Boson_Fermion/IV.A/e60--62,auth:Kimura_Scalar_Fermion/II/e17--18}, and the invertibility of $\mathsf{F}$ is the condition~\citemsp{auth:Kimura_Boson_Fermion/IV.B/e68} for them to be second class. The body therefore has rank $6\mathcal{N}+8$ and, by~\eqref{eq:normal-form}, the consistency of the primary constraints fixes the multipliers of $\tensor{p}{^i_I}$, $\tensor{\chi}{^I_i}$, $\Theta_\Psi$ and $\Theta_{\bar{\Psi}}$, whereas the conditions associated with $\tensor{C}{^\star_I}$ take the form
            \begin{equation}\label{eq:secondary_E}
                \Omega_I + \tensor{\mathcal{S}}{_I_J}\,u^J\approx0\,,\qquad \Omega_I^{(0)} = -N\frac{\partial\mathcal{H}_0^{(0)}}{\partial\tensor{A}{^I_\star}} = -N\left(\tensor{p}{_\phi_I} - \sqrt{\gamma}\,\tensor{h}{_I_J}\tensor{A}{^J_\star}\right),
            \end{equation}
        since at level zero $\tensor{A}{^I_\star}$ only enters $\mathcal{H}_0$ through $\tensor{p}{_\phi_I}\tensor{A}{^I_\star} - \frac{1}{2}\sqrt{\gamma}\,\tensor{h}{_I_J}\tensor{A}{^I_\star}\tensor{A}{^J_\star}$, with $u^J$ the multipliers of $\tensor{C}{^\star_J}$ and $\tensor{{\mathcal{S}^{(0)}}}{_I_J}=0$. Hence, the bodies $u^{J(0)}$ remain undetermined, and~\eqref{eq:secondary_E} defines $\mathcal{N}$ secondary constraints whose body coincides with that of the theory without fermions; at higher levels,~\eqref{eq:secondary_E} relates components determined at lower levels and is solved recursively, as in the superclassical model of~\citemsp{auth:Deffayet_GarciaSaenz/IV.B/e99--104}. Finally, the body of $\{\tensor{C}{^\star_I},\Omega_J\}$ is $-N\sqrt{\gamma}\,\tensor{h}{_I_J}$, which is invertible, so that the consistency of $\Omega_I$ fixes $u^J$ and the algorithm terminates. This realises the possibility left open in~\citemsp{auth:Kimura_Boson_Fermion/AB}, where multipliers that are not determined at level zero, owing to a soul-valued block of the constraint matrix, are fixed by the consistency of the secondary constraints.

        The phase space of $(N,N^i,\tensor{e}{^a_i},\phi^I,\tensor{A}{^I_\star},\tensor{A}{^I_i})$ has dimension $32+10\mathcal{N}$, to which the spinor adds 16 real dimensions. Subtracting twice the 14 first-class constraints $\pi$, $\pi_i$, $\mathcal{H}_\mu$ and $\tensor{\Lambda}{_a_b}$, and the $8\mathcal{N}+8$ second-class constraints $\tensor{p}{^i_I}$, $\tensor{\chi}{^I_i}$, $\tensor{C}{^\star_I}$, $\Omega_I$ and the eight real constraints contained in $\Theta_\Psi$ and $\Theta_{\bar{\Psi}}$, leaves $12+2\mathcal{N}$ dimensions, i.e.\ $2+\mathcal{N}$ bosonic and 4 fermionic degrees of freedom, in agreement with the level decomposition of~\cref{sec:EOM}, which does not rely on the recursive treatment of~\eqref{eq:secondary_E}. The disformal coupling therefore does not introduce an Ostrogradsky ghost, despite the scalar equations containing third time derivatives. Nevertheless, for comparison with previous works, we show in ~\cref{sec:2} a special case with no third time derivatives.

        \subsection{Degeneracy at the level of the equations of motion}
        \label{sec:EOM}

        The same conclusion follows from the field equations. We collect the bosonic fields into $\Xi=(\tensor{e}{^a_\mu},\phi^I)$ and write the action as $S=S_B[\Xi]+S_F[\Xi,\Psi]$, with $S_F$ bilinear in the spinor. The field equations read
            \begin{equation}\label{eq:EOM_split}
                \frac{\delta S_B}{\delta\Xi}+\frac{\delta S_F}{\delta\Xi}=0\,,\qquad \frac{\delta S_F}{\delta\bar{\Psi}}\equiv\mathcal{D}[\Xi]\,\Psi=0\,,
            \end{equation}
        where $\mathcal{D}[\Xi]$ is a first-order operator whose coefficients depend on $\Xi$ and, through $M_\alpha$, on the second derivatives of the $\phi^I$. The term $\delta S_F/\delta\Xi$ is a sum of bilinears and their derivatives multiplied by functions of $\Xi$, and contains the third-order derivatives of~\cref{sec:2}, all of which are proportional to $J_5^\alpha$. Expanding $\Xi=\sum_n\Xi^{(2n)}$ and $\Psi=\sum_n\Psi^{(2n+1)}$ according to~\eqref{eq:level}, and using that $\delta S_F/\delta\Xi$ has no component at level zero, one obtains the hierarchy
            \begin{equation}\label{eq:hierarchy}
                \begin{aligned}
                    &\frac{\delta S_B}{\delta\Xi}\big[\Xi^{(0)}\big]=0\,,\\
                    &\mathcal{D}\big[\Xi^{(0)}\big]\,\Psi^{(2n+1)} = -\sum_{k=1}^n\mathcal{D}^{(2k)}\,\Psi^{(2n+1-2k)}\,,\\
                    &\mathfrak{L}\big[\Xi^{(0)}\big]\cdot\Xi^{(2n)} = \mathcal{Q}^{(2n)}\big[\Xi^{(0)},\ldots,\Xi^{(2n-2)};\Psi^{(1)},\ldots,\Psi^{(2n-1)}\big]\,,\qquad n\ge1\,,
                \end{aligned}
            \end{equation}
        where $\mathfrak{L}[\Xi^{(0)}]$ is the linearisation of $\delta S_B/\delta\Xi$ around the body, $\mathcal{D}^{(2k)}$ is the level-$2k$ component of $\mathcal{D}[\Xi]$, and the source $\mathcal{Q}^{(2n)}$ collects the remaining terms, including all fermionic contributions. At level zero the fermion drops out and one recovers the bosonic theory. At every higher level the equations are linear in the new unknown, with the principal part of the bosonic theory, or of the Dirac operator, on the body background, while the higher derivatives generated by the fermionic coupling only act on lower-level components, which are already determined. The integrability conditions of these linear systems, i.e.\ the linearised Bianchi identities, are satisfied, since the Noether identities of the full action hold for supernumber-valued fields at each level. Hence, each level requires the same initial data as the system without coupling, irrespective of the order of the full equations. This holds also if the bosonic sector is degenerate, as its linearisation inherits the degeneracy.

        This is illustrated by the toy model
            \begin{equation}\label{eq:toy}
                \ddot{q} + V^\prime(q) + \bar{\Psi} \Psi\, \dddot{q} =0\,,
            \end{equation}
        with $q$ Grassmann even and $\bar{\Psi}\Psi$ a given bilinear, which mimics the structure of the scalar field equations, cf.~\eqref{eq:third-order}. Expanding in levels,
            \begin{equation}
                \ddot{q}^{(0)} + V^\prime(q^{(0)}) = 0\,, \qquad
                \ddot{q}^{(2)} + V^{\prime\prime}(q^{(0)})\, q^{(2)} = -\big(\bar{\Psi}\Psi\big)^{(2)}\dddot{q}^{(0)} = \big(\bar{\Psi}\Psi\big)^{(2)}V^{\prime\prime}(q^{(0)})\,\dot{q}^{(0)}\,,\quad\ldots
            \end{equation}
        so that two initial conditions fix $q(t)$ at every level. The third derivative only acts on lower-level components, where it can be eliminated using the time derivative of the lower-level equation; this is how the higher derivatives of the full system can be removed by combining the field equations. Similarly, in~\citemsp{auth:Kimura_Boson_Fermion/IV.C/e81--82,auth:Kimura_Scalar_Fermion/V.A} the second derivatives of the fermions are removed from the bosonic equations using the time derivative of the fermionic ones. A similar discussion of a superclassical model with a higher-derivative coupling is given in~\citemsp{auth:Deffayet_GarciaSaenz/IV.B/e83--88}.

        \subsection{Hamiltonian analysis in the Jordan frame}
        \label{subsec:Hamiltonian}

        For single-field quadratic DHOST theories it has been shown in~\citemsp{auth:Deffayet_GarciaSaenz/IV.A} that a minimally coupled fermion does not spoil the degeneracy conditions. This conclusion does not depend on the number of degrees of freedom or the structure of the constraints, but solely on the distinction between body and soul~\citemsp{auth:Deffayet_GarciaSaenz/IV.B}, and therefore carries over to multi-field DHOST theories, as we now show. Whenever the transformation is invertible, the result is the Jordan-frame counterpart of~\cref{sec:Einstein_frame}.

        We use the ADM variables of~\cref{sec:Einstein_frame} for the Jordan-frame metric, dropping the tildes. Following~\citemsp{BouzariNezhad:2026zsj/4.1/e4.1}, the kinetic part of a generic multi-field quadratic DHOST action reads
            \begin{equation}\label{eq:Lkin}
                \mathcal{L}_{\text{kin}} = \frac{1}{2} \tensor{\mathcal{K}}{^i^j^k^l} \tensor{K}{_i_j} \tensor{K}{_k_l} + \tensor{\mathcal{K}}{^i^j_I} \tensor{K}{_i_j} \tensor{V}{^I_\star} + \frac{1}{2} \tensor{\mathcal{M}}{_I_J} \tensor{V}{^I_\star} \tensor{V}{^J_\star}\,,
            \end{equation}
        where $\tensor{V}{^I_\star} = n^\mu n^\nu \tensor{\phi}{^I_\mu_\nu}$, given by~\eqref{eq:3+1} after the introduction of the auxiliary fields, carries the second time derivative of $\phi^I$, and $\tensor{\mathcal{K}}{^i^j^k^l}$, $\tensor{\mathcal{K}}{^i^j_I}$ and $\tensor{\mathcal{M}}{_I_J}$ are the blocks of the kinetic matrix on $(\tensor{K}{_i_j}, \tensor{V}{^I_\star})$, fixed by the DHOST free functions. The DHOST degeneracy conditions~\citemsp{BouzariNezhad:2026zsj/5.4/e5.12} imply the existence of $\mathcal{N}$ independent null directions of the kinetic matrix; for invertible $\tensor{\mathcal{K}}{^i^j^k^l}$ they read $\tensor{\mathcal{M}}{_I_J}=\tensor{\mathcal{K}}{^i^j_I}\tensor{(\mathcal{K}^{-1})}{_i_j_k_l}\tensor{\mathcal{K}}{^k^l_J}$. As a consequence, the theory possesses $\mathcal{N}$ primary constraints of the form~\citemsp{BouzariNezhad:2026zsj/6.2/e6.17}
            \begin{align}\label{eq:PhiI}
               \Phi_I \equiv \tensor{p}{_\star_I} - 2 \tensor{W}{_I_i_j} \tensor{\pi}{^i^j} - \Delta_I \approx 0\,,
            \end{align}
        where $\tensor{p}{_\star_I}$ and $\tensor{\pi}{^i^j}$ denote the canonical momenta conjugate to $\tensor{A}{^I_\star}$ and $\tensor{\gamma}{_i_j}$ respectively, $\tensor{W}{_I_i_j}=\tensor{(\mathcal{K}^{-1})}{_i_j_k_l}\tensor{\mathcal{K}}{^k^l_I}$ is set by the null eigenvectors, and $\Delta_I$ contains terms independent of the velocities. These constraints are those responsible for the degeneracy that removes the potentially harmful Ostrogradsky modes. In the absence of fermions, their consistency generates $\mathcal{N}$ secondary constraints $\Omega_I$, provided the multi-field consistency conditions~\citemsp{BouzariNezhad:2026zsj/6.5/e6.52}, which guarantee that the consistency of $\Phi_I$ does not fix their multipliers, and the rank condition~\citemsp{BouzariNezhad:2026zsj/6.6/e6.62}, which guarantees that $(\Phi_I,\Omega_I)$ are second class, hold.

        We now minimally couple the Dirac spinor~\eqref{eq:dirac-matter}. In the tetrad formulation $\tensor{\pi}{^i^j}\equiv\frac{1}{2}\tensor{e}{^a^{(i}}\tensor{\pi}{_a^{j)}}$, where $\tensor{\pi}{_a^i}$ is conjugate to $\tensor{e}{^a_i}$. By~\eqref{eq:twist}, the velocities of the tetrad enter the spinor action only through the totally antisymmetric part of the spin connection, i.e.\ through the rotation of the triad, so that the spinor contributes only to the antisymmetric part of $\tensor{\pi}{_a^i}$. Consequently, $\tensor{\pi}{^i^j}$ and $\tensor{p}{_\star_I}$ are unaffected and the constraints~\eqref{eq:PhiI} retain their bosonic form (cf.~\citemsp{auth:Deffayet_GarciaSaenz/IV.A/e73,auth:Takahashi_Matter_Coupling/III.B/e44}). The spinor contributes the spin density to the Lorentz generator $\tensor{\Lambda}{_a_b}\approx0$, which remains first class, and the fermionic constraints of~\eqref{eq:primary_E} for $\mathcal{U}=\mathds{1}$~\citemsp{auth:Deffayet_GarciaSaenz/IV.A/e76},
           \begin{equation}\label{eq:Theta_J}
               \Theta_\Psi\equiv\pi_\Psi-\frac{1}{2}\sqrt{\gamma}\,\bar{\Psi} E^0\approx0\,,\qquad \Theta_{\bar{\Psi}}\equiv\pi_{\bar{\Psi}}-\frac{1}{2}\sqrt{\gamma}\,E^0\Psi\approx0\,,\qquad E^0=-n_\mu\gamma^\mu\,,
           \end{equation}
        where $E^0$ is the spinor-space matrix from the time component of the tetrad. Since $(E^0)^2=-\mathds{1}$, the bracket $\{\Theta_\Psi,\Theta_{\bar{\Psi}}\}=-\sqrt{\gamma}\,(E^0)^{\mathrm{T}}$ is invertible and renders $\Theta_\Psi$, $\Theta_{\bar{\Psi}}$ second class~\citemsp{auth:Deffayet_GarciaSaenz/IV.A/e81}.

        Apart from $\tensor{p}{^i_I}\approx0$ and $\tensor{\chi}{^I_i}\approx0$, which form second-class pairs as in~\cref{sec:Einstein_frame} and play no role in the argument, the primary constraints which are not first class are $C_{\mathsf{M}} = (\Phi_I,\Theta_\Psi,\Theta_{\bar{\Psi}})$, with constraint matrix $\tensor{\mathsf{C}}{_{\mathsf{M}}_{\mathsf{N}}} = \{ C_{\mathsf{M}} , C_{\mathsf{N}} \}$. The bracket $\{\Phi_I,\Phi_J\}$ is that of the bosonic theory. We note that in general the brackets $\{\Phi_I,\Theta_\Psi\}$ and $\{\Phi_I,\Theta_{\bar{\Psi}}\}$ are non-vanishing, since $\Phi_I$ contains $\tensor{\pi}{^i^j}$, which does not commute with $\sqrt{\gamma}$, and their explicit form depends on the multi-field DHOST model of our choosing. Nevertheless, this does not influence the present argument. Since the brackets are linear in the spinor, they are Grassmann odd and contribute only to the soul of the constraint matrix. The body is given by (cf.~\citemsp{auth:Deffayet_GarciaSaenz/IV.A/e82})
            \begin{equation}\label{eq:body_DHOST}
                \mathsf{C}^{(0)} = 
                    \begin{pmatrix}
                        \{\Phi_I,\Phi_J\}^{(0)} & 0 \\
                        0 & \mathsf{F}_0
                    \end{pmatrix},\qquad
                \mathsf{F}_0 = -\sqrt{\gamma}\begin{pmatrix}0 & (E^0)^{\mathrm{T}}\\ E^0 & 0\end{pmatrix}.
            \end{equation}
        The lower right block is invertible, while the upper left block vanishes weakly by the consistency conditions~\citemsp{BouzariNezhad:2026zsj/6.5/e6.52}. By~\eqref{eq:normal-form}, the consistency of the primary constraints fixes the multipliers of $\Theta_\Psi$ and $\Theta_{\bar{\Psi}}$, while the remaining $\mathcal{N}$ conditions are of the form~\eqref{eq:secondary_E}, with $\Omega_I^{(0)}$ the secondary constraints of the bosonic theory and the soul corrections determined recursively. Since nilpotent corrections cannot alter the rank of a body, the rank condition~\citemsp{BouzariNezhad:2026zsj/6.6/e6.62} guarantees that $(\Phi_I,\Omega_I)$ remain second class. As such, the fermionic sector contributes only the standard second-class fermionic constraints and provides no mechanism for reintroducing Ostrogradsky modes. Consequently, the full system propagates the $2+\mathcal{N}$ degrees of freedom of the bosonic multi-field DHOST theory and the 4 degrees of freedom of the Dirac spinor.

    \section{Disformal transformations with higher derivatives}
    \label{sec:higher_derivative}

        Let us return to a single scalar field and consider the disformal transformation~\citemsp{auth:Takahashi_Matter_Coupling/II.A/e8,Takahashi:2022mew}
            \begin{align}\label{eq:GDH}
                \tensor{\tilde{g}}{_\mu_\nu} = f_0\, \tensor{g}{_\mu_\nu} + f_1\, \phi_\mu \phi_\nu + 2f_2\, \phi_{(\mu} \mathcal{X}_{\nu)} + f_3\, \mathcal{X}_\mu \mathcal{X}_\nu\,, \qquad \mathcal{X}_\mu = \left(\tensor{\delta}{^\alpha_\mu} - \frac{\phi^\alpha \phi_\mu}{X}\right) \partial_\alpha X\,,
            \end{align}
        with $f_i=f_i(\phi,X,\mathcal{Z})$ and $\mathcal{Z}\equiv\mathcal{X}_\mu\mathcal{X}^\mu$, subject to the invertibility conditions of~\citemsp{auth:Takahashi_Matter_Coupling/II.A/e19}. Compared with the general invertible transformations of~\cite{auth:Takahashi_Invertible_Disformal}, which depend on $\partial_\mu X$ also through $\phi^\mu\partial_\mu X$, the projection in $\mathcal{X}_\mu$ removes the component along $\phi^\mu$. In the unitary gauge, $\phi=\phi(t)$, one has $\phi_\mu=-\sqrt{-X}\,n_\mu$ for $\dot{\phi}>0$, with $X=-\dot{\phi}^2/N^2$, so that $\phi^\mu\partial_\mu X$ contains $\dot{N}$, while $\mathcal{X}_\mu$ only retains the spatial gradient of $X$, $\mathcal{X}_\mu=-2X a_\mu$, with $a_\mu=D_\mu\ln N$ the acceleration of the normal. The transformation~\eqref{eq:GDH} then reads
            \begin{align}\label{eq:GDH_unitary}
                \tensor{\tilde{g}}{_\mu_\nu} = f_0\, \tensor{g}{_\mu_\nu} - f_1 X\, n_\mu n_\nu + 4f_2\, X\sqrt{-X}\, n_{(\mu}a_{\nu)} + 4f_3 X^2\, a_\mu a_\nu\,, \qquad \mathcal{Z}=4X^2\,a_ia^i\,,
            \end{align}
        so that the lapse, shift and spatial metric of $\tilde{\mathbf{g}}$ are functions of $N$, $N^i$, $\tensor{\gamma}{_i_j}$ and $D_iN$, but not of $\dot{N}$. Considering only bosons, this leads to a theory free of the Ostrogradsky ghost in the unitary gauge~\cite{Takahashi:2022mew,Naruko:2022vuh}, while away from it a shadowy mode appears~\cite{DeFelice:2018ewo,DeFelice:2021hps}. Therefore, inserting the inverse of~\eqref{eq:GDH} into a Horndeski action leads, in the absence of fermions, to a degenerate theory which, in the unitary gauge, belongs to the spatially covariant gravity theories with the velocity of the lapse function, $F\equiv n^\mu\partial_\mu N$, and satisfies the degeneracy and consistency conditions of~\citemsp{auth:Gao_Yao_SCG/4}.

        In the unitary gauge, time reparametrisations are fixed. Compared with the covariant analysis of~\cref{sec:Einstein_frame}, the two first-class constraints associated with them, namely the vanishing of the momentum conjugate to the lapse and the Hamiltonian constraint, are therefore missing, and the scalar degree of freedom is carried by the lapse. In metric variables, following~\citemsp{auth:Gao_Yao_SCG/2/e2.4}, we replace $F$ and $\tensor{K}{_i_j}$ by auxiliary variables $A$ and $\tensor{B}{_i_j}$, enforced by the Lagrange multipliers $\lambda$ and $\tensor{\lambda}{^i^j}$,
            \begin{align}\label{eq:GY_action}
                S = \int\!\md^4x\,\sqrt{\gamma}\,N\left[\mathcal{L}(t,N,\tensor{\gamma}{_i_j},A,\tensor{B}{_i_j},D_k) + \lambda\left(F-A\right) + \tensor{\lambda}{^i^j}\left(\tensor{B}{_i_j}-\tensor{K}{_i_j}\right)\right],
            \end{align}
        where $\mathcal{L}$ depends on spatial derivatives through $D_k$. The velocities $\dot{N}$ and $\tensor{\dot{\gamma}}{_i_j}$ only enter through the multiplier terms, so that $\pi=\sqrt{\gamma}\,\lambda$ and $\tensor{\pi}{^i^j}=-\frac{1}{2}\sqrt{\gamma}\,\tensor{\lambda}{^i^j}$, while the momenta conjugate to $N^i$, $A$ and $\tensor{B}{_i_j}$ vanish. Eliminating the multipliers through the equations of motion of $A$ and $\tensor{B}{_i_j}$ yields the primary constraints $\pi_i\approx0$, $p\approx0$, $\tensor{p}{^i^j}\approx0$ and~\citemsp{auth:Gao_Yao_SCG/3.1/e3.6}
            \begin{align}\label{eq:GY_primary}
                \tilde{\pi} \equiv \pi - \frac{1}{N} \frac{\delta S_{\mathcal{L}}}{\delta A}\approx0\,,\qquad \tensor{\tilde{\pi}}{^i^j} \equiv \tensor{\pi}{^i^j} - \frac{1}{2 N} \frac{\delta S_{\mathcal{L}}}{\delta \tensor{B}{_i_j}}\approx0\,,
            \end{align}
        where $S_{\mathcal{L}}$ is the first term in~\eqref{eq:GY_action}, and $\pi$, $\pi_i$, $\tensor{\pi}{^i^j}$, $p$ and $\tensor{p}{^i^j}$ are conjugate to $N$, $N^i$, $\tensor{\gamma}{_i_j}$, $A$ and $\tensor{B}{_i_j}$. Only $\pi_i\approx0$ and the momentum constraints $\mathcal{H}_i\approx0$, which generate spatial diffeomorphisms, are first class. The phase space of $(N,N^i,\tensor{\gamma}{_i_j},A,\tensor{B}{_i_j})$ has dimension 34. Subtracting twice the 6 first-class constraints and the 14 constraints $p$, $\tensor{p}{^i^j}$, $\tilde{\pi}$ and $\tensor{\tilde{\pi}}{^i^j}$, which are second class for a non-degenerate kinetic matrix, leaves 8 dimensions, i.e.\ 4 degrees of freedom. The degeneracy and consistency conditions of~\citemsp{auth:Gao_Yao_SCG/4} yield two additional second-class constraints, which remove the fourth one, so that the form of~\eqref{eq:GY_primary} is crucial to obtain just three degrees of freedom.

        In this frame the fermion is minimally coupled, and we can essentially copy the Hamiltonian analysis of~\cref{subsec:Hamiltonian}. One point requires care: the spinor couples directly to the velocity of the tetrad, while~\eqref{eq:GY_primary} involves the momentum conjugate to the spatial metric. Choosing the time gauge $n^a=\tensor{\delta}{^a_{\hat{0}}}$ in~\eqref{eq:tetrad_ADM}, i.e.\ $\theta^{\hat{0}}=N\md t$ and $\theta^{\hat{i}}=\tensor{e}{^{\hat{i}}_k}(\md x^k+N^k\md t)$, where hats denote Lorentz indices, the spin connection contains neither $\dot{N}$ nor $\dot{N}^i$~\citemsp{auth:Takahashi_Matter_Coupling/III.B/e41}, and the triad velocity enters the Hermitian Dirac action only through $\tensor{e}{_{\hat{i}}^k}\tensor{\dot{e}}{_{\hat{j}}_k}\tensor{\gamma}{^{\hat{i}}^{\hat{j}}}$~\citemsp{auth:Takahashi_Matter_Coupling/III.B/e44}, with $\tensor{\gamma}{^{\hat{i}}^{\hat{j}}}\equiv\gamma^{[\hat{i}}\gamma^{\hat{j}]}$, i.e.\ through its antisymmetric part, contracted with $\Psi^\dagger\tensor{\gamma}{^{\hat{i}}^{\hat{j}}}\Psi\propto\tensor{\epsilon}{^{\hat{i}}^{\hat{j}}^{\hat{k}}}J_{5\hat{k}}$, in agreement with~\eqref{eq:twist}. The constraint associated with the triad momentum $\tensor{\pi}{_{\hat{i}}^k}$ therefore reads
            \begin{align}\label{eq:triad_constraint}
                \tensor{\tilde{\pi}}{_{\hat{i}}^k} \equiv \tensor{\pi}{_{\hat{i}}^k} - \frac{1}{N} \tensor{e}{_{\hat{i}}_j} \frac{\delta S_{\mathcal{L}}}{\delta \tensor{B}{_j_k}} - \sqrt{\gamma}\, \tensor{s}{_{\hat{i}}^k}\approx0\,,
            \end{align}
        where $\tensor{s}{_{\hat{i}}^k}$ is bilinear in the spinor and satisfies $\tensor{e}{^{\hat{i}}^{(j}}\tensor{s}{_{\hat{i}}^{k)}}=0$. Its symmetric part, $\frac{1}{2}\tensor{e}{^{\hat{i}}^{(j}}\tensor{\tilde{\pi}}{_{\hat{i}}^{k)}}$, coincides with $\tensor{\tilde{\pi}}{^j^k}$ in~\eqref{eq:GY_primary}, while its antisymmetric part is the generator of spatial rotations, including the spin density, which is first class. The spinor thus only shifts the latter, and the constraints~\eqref{eq:GY_primary}, together with $\pi_i\approx0$, $p\approx0$ and $\tensor{p}{^i^j}\approx0$ for the shift and the auxiliary variables, retain the form of the bosonic theory. Also, we have to include again the fermionic constraints $\Theta_\Psi$ and $\Theta_{\bar{\Psi}}$, with $E^0=\gamma^{\hat{0}}$, whose brackets with the constraints~\eqref{eq:GY_primary} are Grassmann odd. By~\eqref{eq:normal-form}, the body of the constraint matrix, and with it the degeneracy and consistency conditions of~\citemsp{auth:Gao_Yao_SCG/4}, are those of the bosonic theory, so that the additional constraints are not spoiled. In the time gauge, the three additional components of the triad are removed by the first-class rotation generators, and the spinor adds 16 real dimensions and 8 second-class constraints, so that the theory propagates 3 bosonic and 4 fermionic degrees of freedom.

        The same conclusion is reached in the Einstein frame, where the fermion is coupled to~\eqref{eq:GDH}. There, after a rescaling of the spinor, the Dirac action contains the term~\citemsp{auth:Takahashi_Matter_Coupling/III.B/e48}
            \begin{align}\label{eq:TKM_term}
                S_{\mathrm{Dirac}}\supset\int\md^4x\,\sqrt{\gamma}\,\mathfrak{f}\;\partial_kN\,\partial_l\dot{N}\;\Psi^\dagger \tensor{e}{_{\hat{i}}^k}\tensor{e}{_{\hat{j}}^l}\tensor{\gamma}{^{\hat{i}}^{\hat{j}}}\Psi\,,
            \end{align}
        where $\mathfrak{f}\propto E_4^2$, with $E_4$ given in~\citemsp{auth:Takahashi_Matter_Coupling/II.B/e23}, vanishes if and only if $f_3=0$. In~\citemsp{auth:Takahashi_Matter_Coupling/III.B} it was concluded that $\dot{N}$ cannot be removed from the lapse equation without reintroducing $\dot{\Psi}$, which led to the condition $f_3=0$~\citemsp{auth:Takahashi_Matter_Coupling/III.B/e51}, see also~\cite{auth:Ikeda_Consistency_Couplings}, and~\citemsp{auth:Domenech_consistency_conditions/IV.A/e82} for a closely related condition. These arguments treat the bilinear in~\eqref{eq:TKM_term} as an ordinary function. However,~\eqref{eq:TKM_term} is again a coupling to the axial current and hence a soul. The momentum conjugate to the lapse becomes a soul, $\pi-\sqrt{\gamma}\, s_N\approx0$, whose body is the primary constraint $\pi\approx0$ of Horndeski theory in the unitary gauge, and, by~\cref{sec:EOM}, $\dot{N}$ enters the field equations only at level two and higher, multiplied by components determined at lower levels. These conditions are therefore not required: fermions impose no condition beyond those required by bosonic matter.

    \section{Conclusions}
    \label{sec:concl}

        We have extended the analysis of disformal couplings to Dirac fermions to the case of $\mathcal{N}$ scalar fields. The induced axial coupling $M_\alpha$ generically introduces third-order derivatives of the scalar fields $\phi^K$. For field-dependent free functions, manifestly second-order equations of motion in the Einstein frame require $\mathrm{rank}(\tensor{d}{_I_J}) \le 1$, which restricts the disformal transformation to act along a single field-space direction $A^1$, directly generalizing the result previously established for two scalar fields~\citemsp{auth:Domenech_consistency_conditions/III.B/e54} and confirming the conjecture of~\citemsp{auth:Domenech_consistency_conditions/VI}. For $\tensor{X}{^K^L}$-dependent transformations along $A^1$, the derivatives of its direction with respect to $\tensor{X}{^K^L}$ must in addition be parallel. The axial coupling then vanishes if and only if the Frobenius integrability condition $A^1 \wedge \mathrm{d}A^1 = 0$ is satisfied, which, unlike in the two-scalar field setting, is a non-trivial condition for $\mathcal{N}\ge3$.

        Our main result is that these conditions are not required for the absence of Ostrogradsky ghosts. After the transformation, the fermion registers the derivatives of the matter-frame coframe only through the totally antisymmetric part of the spin connection, contracted with the axial current, and the second derivatives of the scalar fields enter only through $M_\alpha J_5^\alpha$. All couplings between the fermion and the scalar fields, including those responsible for the higher derivatives, therefore multiply a fermion bilinear, which is Grassmann even and has vanishing body. The Hamiltonian analysis in the Einstein frame shows that these couplings only contribute to the soul of the constraint matrix, whose body is that of the theory without fermions, so that the primary constraints associated with $\tensor{A}{^I_\star}$ generate the same secondary constraints and the theory propagates $2+\mathcal{N}$ bosonic and 4 fermionic degrees of freedom, for free functions depending on $\phi^K$ and $\tensor{X}{^K^L}$. Equivalently, the level decomposition turns the field equations into a hierarchy of linear equations with the principal part of the bosonic theory, the higher derivatives acting only on lower-level components. This explains why the higher derivatives can be removed by combining the field equations.

        The same mechanism applies in the Jordan frame. For multi-field DHOST theories with minimally coupled fermions, the spinor does not modify the DHOST primary constraints, and their brackets with the fermionic constraints are Grassmann odd, so that the multi-field degeneracy, consistency and rank conditions of~\citemsp{BouzariNezhad:2026zsj/6} are unaffected, generalizing the single-field result of~\cite{auth:Deffayet_GarciaSaenz}. For disformal transformations with higher derivatives of a single scalar field, the spinor only shifts the generator of spatial rotations in the Jordan frame, and the degeneracy and consistency conditions of~\cite{auth:Gao_Yao_SCG} are unaffected. The term containing $\dot{N}$ that appears in the Einstein frame is again a coupling to the axial current, and the conditions on $f_3$ of~\cite{auth:Takahashi_Matter_Coupling,auth:Domenech_consistency_conditions} are not required.

        Our analysis is classical, with Grassmann-valued spinors, which is the framework in which the degrees of freedom of fermions are counted~\citemsp{auth:DeWitt_Supermanifolds,auth:Henneaux_Teitelboim,auth:Kimura_Boson_Fermion/II}. A coarse-grained description in which fermion bilinears acquire c-number values, e.g.\ a spin-polarized medium, falls outside its scope, as do loop corrections, which generate bosonic higher-derivative operators to be treated within effective field theory.

        It remains an open question if multi-field DHOST theories are closed under disformal transformations with $\tensor{X}{^K^L}$-dependent free functions, as is the case for a single scalar field, which is tied to the invertibility of these transformations. Our results show that fermions pose no obstruction to this closure, nor to the definition of multi-field Horndeski theories via disformal transformations proposed in~\cite{Katayama:2025hnd}. Looking ahead, it would be particularly interesting to examine the direct impact of these disformal transformations on the multi-field degeneracy conditions recently derived in~\citemsp{BouzariNezhad:2026zsj/5--6}.

\begin{acknowledgments}
We would like to thank G. Dom\`enech and K. Takahashi for their helpful comments and discussion. This work was partially supported by the DFG under the Emmy-Noether program, project number 496592360. 
\end{acknowledgments}

\appendix

    \section{Second-order equations in the Einstein frame}
    \label{sec:2}

        In this appendix we derive the conditions under which the scalar field equations in the Einstein frame are manifestly second order. We write the axial term of~\eqref{eq:dirac-gravity} as
            \begin{equation}\label{eq:axial-term}
                S_{\mathrm{Dirac}}\supset -i\int\!\md^4x\,\sqrt{-g}\;\mathcal{W}\,M_\alpha J_5^\alpha\,,\qquad \mathcal{W}\equiv C^3\det\mathcal{U}\,,
            \end{equation}
        and treat $J_5^\alpha$ as given, since the conditions only concern the derivatives of the scalar fields. For $\mathcal{N}$ fields the axial coupling reads
            \begin{align}\label{eq:M-full}
                M^\alpha = & \frac{1}{4} \tensor{\epsilon}{^\alpha^\beta^\gamma^\delta} \tensor{d}{_K_L} \tensor{\phi}{^L_\gamma} \tensor{\phi}{^I_\beta} \Big[ \tensor{X}{^K^J} \nabla_\delta \tensor{c}{_I_J} + \tensor{c}{_I_J} \tensor{\phi}{^K^\nu} \tensor{\phi}{^J_\nu_\delta} - \tensor{d}{_M_I} \tensor{c}{_Q_J} \tensor{X}{^K^J} \tensor{\phi}{^M^\mu} \tensor{\phi}{^Q_\mu_\delta} \nonumber \\
                & + \tensor{d}{_M_Q} \tensor{\phi}{^M^\mu} \tensor{\phi}{^Q_\delta} \left( \tensor{X}{^K^J} \nabla_\mu \tensor{c}{_I_J} + \tensor{c}{_I_J} \tensor{\phi}{^K^\nu} \tensor{\phi}{^J_\nu_\mu} \right) \Big] \nonumber \\
                = & \frac{1}{4} \tensor{\epsilon}{^\alpha^\beta^\gamma^\delta} \tensor{\phi}{^L_\gamma} \tensor{\phi}{^I_\beta}  \Big[ \tensor{T}{_R_L_I_S}\, \tensor{\phi}{^R^\nu} \tensor{\phi}{^S_\nu_\delta}  + \tensor{d}{_R_L} \tensor{d}{_M_Q} \tensor{c}{_I_S}  \tensor{\phi}{^Q_\delta} \tensor{\phi}{^M^\mu} \tensor{\phi}{^R^\nu} \tensor{\phi}{^S_\nu_\mu} \nonumber \\
                & + \tensor{d}{_K_L} \tensor{X}{^K^J}  \nabla_\delta \tensor{c}{_I_J} + \tensor{d}{_K_L} \tensor{d}{_M_Q} \tensor{X}{^K^J} \tensor{\phi}{^M^\mu} \tensor{\phi}{^Q_\delta} \nabla_\mu \tensor{c}{_I_J}   \Big]\,,
            \end{align}
        with $\tensor{T}{_R_L_I_S}\equiv \tensor{d}{_R_L} \tensor{c}{_I_S} - \tensor{d}{_K_L}  \tensor{d}{_R_I} \tensor{c}{_S_J} \tensor{X}{^K^J}$. For one scalar field the axial coupling vanishes completely, while for two we recover the results in~\citemsp{auth:Domenech_consistency_conditions/III.A/e27--28}. Since $M^\alpha$ is affine in $\tensor{\phi}{^S_\mu_\nu}$, the third derivatives follow from the fact that a term $\sqrt{-g}\,\tensor{\mathfrak{F}}{^\mu^\nu_S}(\phi^K,\tensor{\phi}{^K_\rho})\,\tensor{\phi}{^S_\mu_\nu}$ in the Lagrangian contributes
            \begin{equation}\label{eq:F-third}
                \frac{1}{\sqrt{-g}}\frac{\delta S}{\delta\phi^B}\supset\left(\frac{\partial\tensor{\mathfrak{F}}{^\mu^\nu_B}}{\partial\tensor{\phi}{^C_\lambda}}-\frac{\partial\tensor{\mathfrak{F}}{^\mu^\nu_C}}{\partial\tensor{\phi}{^B_\lambda}}\right)\nabla_\lambda\nabla_\mu\nabla_\nu\phi^C
            \end{equation}
        to the equation of motion of $\phi^B$, up to terms with at most second derivatives arising from the commutation of covariant derivatives.

        \subsection{Field-dependent transformations}
        \label{subsec:field_dependent}

            Let $\tensor{c}{_I_J}=\tensor{c}{_I_J}(\phi^K)$, so that the last two terms in~\eqref{eq:M-full} contain no second derivatives. The derivatives in~\eqref{eq:F-third} acting on $\tensor{\phi}{^R^\nu}$ in the first term produce the structure $\tensor{\epsilon}{^\alpha^\beta^\gamma^\delta}\nabla_\delta\nabla^\nu\nabla_\nu\phi^C$, which no other term produces: derivatives acting on the gradients contracted with the Levi-Civita symbol place a second index of the symmetric third derivative on it in the first term, and two indices on gradients in the second, while derivatives of the coefficients yield gradients contracted with the third derivative. Since $\tensor{\epsilon}{^\alpha^\beta^\gamma^\delta}\tensor{\phi}{^L_\gamma}\tensor{\phi}{^I_\beta}$ is antisymmetric in $(L,I)$, and~\eqref{eq:Inverse_relation} implies $\tensor{d}{_K_L}\tensor{X}{^K^J}\tensor{c}{_J_S}=-\tensor{(c+d)}{_L_S}$, only $\tensor{d}{_R_{[I}}\tensor{d}{_{L]}_S}$ survives in $\tensor{T}{_R_L_I_S}$, and~\eqref{eq:F-third} gives
                \begin{equation}\label{eq:third-order}
                    \frac{1}{\sqrt{-g}}\frac{\delta S}{\delta\phi^B}\supset -\frac{i}{2}\,\mathcal{W}\,J_{5\alpha}\,\tensor{\epsilon}{^\alpha^\beta^\gamma^\delta}\tensor{\phi}{^L_\gamma}\tensor{\phi}{^I_\beta}\,\tensor{d}{_L_B}\,\tensor{d}{_I_C}\,\nabla_\delta\nabla^\nu\nabla_\nu\phi^C\,.
                \end{equation}
            Second-order equations therefore require
                \begin{equation}\label{eq:DTD_components}
                    \phi^{[I}\phi^{L]}\,\tensor{d}{_I_C}\,\tensor{d}{_L_B}=0\,,
                \end{equation}
            with $\phi^{[I}\phi^{L]}$ evaluated for any pair of spacetime components. For $\mathcal{N}=2$ this is the analogue of the condition $h_2=h_3$ derived in~\citemsp{auth:Domenech_consistency_conditions/III.B/e41}. Fixing two spacetime directions $(\gamma_0,\beta_0)$ and defining the field-space vectors $x^I\equiv\tensor{\phi}{^I_{\gamma_0}}$ and $y^I\equiv\tensor{\phi}{^I_{\beta_0}}$, which are arbitrary, condition~\eqref{eq:DTD_components} reads
                \begin{equation}\label{eq:DTD}
                    \mathbf{D}\left(xy^{\mathrm{T}}-yx^{\mathrm{T}}\right)\mathbf{D}=(\mathbf{D}x)(\mathbf{D}y)^{\mathrm{T}}-(\mathbf{D}y)(\mathbf{D}x)^{\mathrm{T}}=0\qquad\forall\,x,y\in\mathbb{R}^{\mathcal{N}}\,,
                \end{equation}
            where $\mathbf{D}=\mathbf{D}^{\mathrm{T}}$ is the matrix representation of $\tensor{d}{_I_J}$. Since $pq^{\mathrm{T}}-qp^{\mathrm{T}}=0$ if and only if $p\parallel q$, this is equivalent to $\mathbf{D}x\parallel\mathbf{D}y$ for all $x$ and $y$. Fixing $y_0$ with $\mathbf{D}y_0\neq0$, every column of $\mathbf{D}$ lies along $\mathbf{D}y_0$, so that $\mathrm{rank}(\tensor{d}{_I_J})\le1$, and~\eqref{eq:Inverse_relation} implies
                \begin{equation}\label{eq:rank1form}
                    \tensor{d}{_I_J}=\lambda_d\,n_In_J\,,\qquad \tensor{c}{_I_J}=\lambda_c\,n_In_J\,,\qquad \lambda_c=-\frac{\lambda_d}{1+\lambda_d\,(A^1)^2}\,,
                \end{equation}
            with $n_J$ the unit eigenvector, $A^1\equiv n_J\,\md\phi^J$ and $(A^1)^2\equiv \tensor{g}{^\mu^\nu}\tensor{A}{^1_\mu} \tensor{A}{^1_\nu}$. The disformal transformation thus collapses to a single field-space direction, generalizing the two-field result of~\citemsp{auth:Domenech_consistency_conditions/III.B/e55},
                \begin{equation}
                    \mathbf{L}=\lambda_c\,A^1\otimes A^1\,,\qquad \tensor{G}{_I_J}\,\md\phi^I\otimes\md\phi^J=\left(2\lambda_c+\lambda_c^2\,(A^1)^2\right)A^1\otimes A^1\,,
                \end{equation}
            and the axial coupling reduces to
                \begin{equation}\label{eq:Mreduced}
                    M^\alpha = \frac{1}{4}\,\lambda_c\lambda_d\,(A^1)^2\,\tensor{\epsilon}{^\alpha^\beta^\gamma^\delta}\tensor{A}{^1_\gamma}\,\nabla_\delta \tensor{A}{^1_\beta}\;\propto\;\big[\star(A^1\wedge\md A^1)\big]^\alpha\,,
                \end{equation}
            since only $\nabla_{[\delta}\tensor{A}{^1_{\beta]}}=\tfrac{1}{2}\tensor{(\md A^1)}{_\delta_\beta}$ survives the contraction. For $\tensor{c}{_I_J}=\tensor{c}{_I_J}(\phi^K)$, the eigenvector $n_J$ depends on $\phi^K$ only, so that $\md A^1=\partial_K n_J\,\md\phi^K\wedge\md\phi^J$ contains no second derivatives and the scalar equations are second order. Hence, for field-dependent transformations the scalar equations are second order if and only if $\mathrm{rank}(\tensor{d}{_I_J})\le1$. Moreover, since $\lambda_c\lambda_d(A^1)^2\neq0$ for a non-trivial transformation,
                \begin{equation}
                    M^\alpha = 0 \iff A^1\wedge \md A^1 = 0\,,
                \end{equation}
            which is the \emph{Frobenius integrability} condition: $A^1$ admits an integrating factor, $A^1=\varrho\,\md\sigma$. This is \emph{weaker} than closedness, $\md A^1=0$. For $\mathcal{N}=2$ the 3-form $A^1\wedge\md A^1$ is built from $\md\phi^1$ and $\md\phi^2$ only and vanishes identically, so that the condition is non-trivial only for $\mathcal{N}\ge3$.

        \subsection{Transformations depending on \texorpdfstring{$\tensor{X}{^K^L}$}{the kinetic matrix}}
        \label{subsec:X_dependent}

            For $\tensor{c}{_I_J}$ depending on $\tensor{X}{^K^L}$, the terms with $\nabla_\delta \tensor{c}{_I_J}$ in~\eqref{eq:M-full} contribute to the coefficient of $\tensor{\phi}{^R^\nu}\tensor{\phi}{^S_\nu_\delta}$, and~\eqref{eq:DTD_components} turns into a differential condition on $\tensor{c}{_I_J}$. We therefore restrict to transformations along a single direction, $\tensor{c}{_I_J}=\lambda_c\,n_In_J$, with $\lambda_c$ and the unit vector $n_J$ depending on $\phi^K$ and $\tensor{X}{^K^L}$. The reduction~\eqref{eq:Mreduced} still holds, and
                \begin{equation}
                    \mathcal{W} M^\alpha = \frac{\mathcal{W}}{4}\,\lambda_c\lambda_d\,(A^1)^2\,\tensor{\epsilon}{^\alpha^\beta^\gamma^\delta}\left(n_M\,\tensor{n}{_J_{,R}_S}\,\tensor{\phi}{^M_\gamma}\tensor{\phi}{^J_\beta}\,\nabla_\delta \tensor{X}{^R^S} + n_M\,\tensor{n}{_J_{,K}}\,\tensor{\phi}{^M_\gamma}\tensor{\phi}{^J_\beta}\tensor{\phi}{^K_\delta}\right),
                \end{equation}
            with $\tensor{n}{_J_{,K}}\equiv\partial n_J/\partial\phi^K$ and $\tensor{n}{_J_{,R}_S}\equiv\partial n_J/\partial \tensor{X}{^R^S}$. We remove the prefactor by the rescaling
                \begin{equation}\label{eq:rescale-W}
                    \Psi\to\left[\frac{\mathcal{W}}{4}\,\lambda_c\lambda_d\,(A^1)^2\right]^{-1/2}\Psi\,,
                \end{equation}
            and similarly for $\bar{\Psi}$. In the Hermitian formulation no derivatives of the rescaling factor appear, so that the kinetic and mass terms only acquire a factor depending on the first derivatives of the scalar fields, and second derivatives are contained in the first term above only. In~\eqref{eq:F-third}, with $\nabla_\delta \tensor{X}{^R^S}=2\tensor{\phi}{^{(R|}^\nu}\tensor{\phi}{^{|S)}_\nu_\delta}$, the derivatives acting on $\tensor{\phi}{^R^\nu}$ cancel by the symmetry of $\tensor{n}{_J_{,R}_S}$, those acting on the gradients contracted with the Levi-Civita symbol vanish, and the second derivatives of $n_J$ cancel between the two terms, so that
                \begin{equation}\label{eq:third-order-X}
                    \frac{1}{\sqrt{-g}}\frac{\delta S}{\delta\phi^B}\supset -4i\,J_{5\alpha}\,\tensor{\epsilon}{^\alpha^\beta^\gamma^\delta}\tensor{\phi}{^M_\gamma}\tensor{\phi}{^J_\beta}\left(\tensor{n}{_M_{,C}_P}\,\tensor{n}{_J_{,R}_B}-\tensor{n}{_M_{,B}_P}\,\tensor{n}{_J_{,R}_C}\right)\tensor{\phi}{^P^\lambda}\tensor{\phi}{^R^\nu}\,\nabla_\lambda\nabla_\nu\nabla_\delta\phi^C\,.
                \end{equation}
            Since the Levi-Civita symbol only retains the part antisymmetric in $(M,J)$, these terms vanish for arbitrary configurations if and only if $\tensor{n}{_{[M|}_{,O}_P}\,\tensor{n}{_{|J]}_{,R}_S}=0$, i.e.\ if all derivatives $\partial n_J/\partial \tensor{X}{^O^P}$ are parallel to a single field-space vector. This holds for $n_J=n_J(\phi^K)$, and identically for $\mathcal{N}=2$, since the derivatives of the unit vector $n_J$ are orthogonal to it. The term with $\tensor{n}{_J_{,K}}$ does not produce higher derivatives\footnote{Note that it vanishes trivially for two scalar fields.}.

            These conditions ensure that the scalar equations are manifestly second order in the Einstein frame. They are not necessary for the absence of ghosts: the third-order terms~\eqref{eq:third-order} and~\eqref{eq:third-order-X} are proportional to $J_5^\alpha$ and, as shown in~\cref{sec:Einstein_frame,sec:EOM}, do not propagate additional degrees of freedom.

\printbibliography

@article{Kobayashi:2019hrl,
    author = "Kobayashi, Tsutomu",
    title = "{Horndeski theory and beyond: a review}",
    eprint = "1901.07183",
    archivePrefix = "arXiv",
    primaryClass = "gr-qc",
    reportNumber = "RUP-19-3",
    doi = "10.1088/1361-6633/ab2429",
    journal = "Rept. Prog. Phys.",
    volume = "82",
    number = "8",
    pages = "086901",
    year = "2019"
}

@article{auth:BekensteinOriginalDisformal,
    author = "Bekenstein, Jacob D.",
    title = "{The Relation between physical and gravitational geometry}",
    eprint = "gr-qc/9211017",
    archivePrefix = "arXiv",
    reportNumber = "UCSB-TH-92-41, UCSBTH-92-41",
    doi = "10.1103/PhysRevD.48.3641",
    journal = "Phys. Rev. D",
    volume = "48",
    pages = "3641--3647",
    year = "1993"
}

@article{Tsujikawa:2010zza,
    author = "Tsujikawa, Shinji",
    title = "{Modified gravity models of dark energy}",
    eprint = "1101.0191",
    archivePrefix = "arXiv",
    primaryClass = "gr-qc",
    doi = "10.1007/978-3-642-10598-2_3",
    journal = "Lect. Notes Phys.",
    volume = "800",
    pages = "99--145",
    year = "2010"
}

@article{auth:Takahashi_Matter_Coupling,
    author = "Takahashi, Kazufumi and Kimura, Rampei and Motohashi, Hayato",
    title = "{Consistency of matter coupling in modified gravity}",
    eprint = "2212.13391",
    archivePrefix = "arXiv",
    primaryClass = "gr-qc",
    reportNumber = "YITP-22-134",
    doi = "10.1103/PhysRevD.107.044018",
    journal = "Phys. Rev. D",
    volume = "107",
    number = "4",
    pages = "044018",
    year = "2023"
}

@article{auth:Takahashi_Invertible_Disformal,
    author = "Takahashi, Kazufumi and Motohashi, Hayato and Minamitsuji, Masato",
    title = "{Invertible disformal transformations with higher derivatives}",
    eprint = "2111.11634",
    archivePrefix = "arXiv",
    primaryClass = "gr-qc",
    reportNumber = "YITP-21-137",
    doi = "10.1103/PhysRevD.105.024015",
    journal = "Phys. Rev. D",
    volume = "105",
    number = "2",
    pages = "024015",
    year = "2022"
}

@article{auth:Ikeda_Consistency_Couplings,
    author = "Ikeda, Tact and Takahashi, Kazufumi and Kobayashi, Tsutomu",
    title = "{Consistency of higher-derivative couplings to matter fields in scalar-tensor gravity}",
    eprint = "2302.03418",
    archivePrefix = "arXiv",
    primaryClass = "gr-qc",
    reportNumber = "RUP-23-2, YITP-23-11",
    doi = "10.1103/PhysRevD.108.044006",
    journal = "Phys. Rev. D",
    volume = "108",
    number = "4",
    pages = "044006",
    year = "2023"
}

@article{Takahashi:2023vva,
    author = "Takahashi, Kazufumi",
    title = "{Invertible disformal transformations with arbitrary higher-order derivatives}",
    eprint = "2307.08814",
    archivePrefix = "arXiv",
    primaryClass = "gr-qc",
    reportNumber = "YITP-23-91",
    doi = "10.1103/PhysRevD.108.084031",
    journal = "Phys. Rev. D",
    volume = "108",
    number = "8",
    pages = "084031",
    year = "2023"
}

@article{Takahashi:2022mew,
    author = "Takahashi, Kazufumi and Minamitsuji, Masato and Motohashi, Hayato",
    title = "{Generalized disformal Horndeski theories: Cosmological perturbations and consistent matter coupling}",
    eprint = "2209.02176",
    archivePrefix = "arXiv",
    primaryClass = "gr-qc",
    reportNumber = "YITP-22-89",
    doi = "10.1093/ptep/ptac161",
    journal = "PTEP",
    volume = "2023",
    number = "1",
    pages = "013E01",
    year = "2023"
}

@article{Brax:2012hm,
    author = "Brax, Philippe",
    title = "{Lorentz Invariance Violation in Modified Gravity}",
    eprint = "1202.0740",
    archivePrefix = "arXiv",
    primaryClass = "hep-ph",
    doi = "10.1016/j.physletb.2012.05.005",
    journal = "Phys. Lett. B",
    volume = "712",
    pages = "155--160",
    year = "2012"
}

@article{Woodard:2015zca,
    author = "Woodard, Richard P.",
    title = "{Ostrogradsky's theorem on Hamiltonian instability}",
    eprint = "1506.02210",
    archivePrefix = "arXiv",
    primaryClass = "hep-th",
    reportNumber = "UFIFT-QG-15-03",
    doi = "10.4249/scholarpedia.32243",
    journal = "Scholarpedia",
    volume = "10",
    number = "8",
    pages = "32243",
    year = "2015"
}

@article{Ganz:2020skf,
    author = "Ganz, Alexander and Noui, Karim",
    title = "{Reconsidering the Ostrogradsky theorem: Higher-derivatives Lagrangians, Ghosts and Degeneracy}",
    eprint = "2007.01063",
    archivePrefix = "arXiv",
    primaryClass = "hep-th",
    doi = "10.1088/1361-6382/abe31d",
    journal = "Class. Quant. Grav.",
    volume = "38",
    number = "7",
    pages = "075005",
    year = "2021"
}

@article{Domenech:2015tca,
    author = "Dom\`enech, Guillem and Mukohyama, Shinji and Namba, Ryo and Naruko, Atsushi and Saitou, Rio and Watanabe, Yota",
    title = "{Derivative-dependent metric transformation and physical degrees of freedom}",
    eprint = "1507.05390",
    archivePrefix = "arXiv",
    primaryClass = "hep-th",
    doi = "10.1103/PhysRevD.92.084027",
    journal = "Phys. Rev. D",
    volume = "92",
    number = "8",
    pages = "084027",
    year = "2015"
}

@article{Horndeski:2024hee,
    author = "Horndeski, Gregory W.",
    title = "{Second-Order Bi-Scalar-Tensor Field Equations in a Space of Four-Dimensions}",
    eprint = "2405.08303",
    archivePrefix = "arXiv",
    primaryClass = "gr-qc",
    month = "5",
    year = "2024"
}

@article{Ohashi:2015fma,
    author = "Ohashi, Seiju and Tanahashi, Norihiro and Kobayashi, Tsutomu and Yamaguchi, Masahide",
    title = "{The most general second-order field equations of bi-scalar-tensor theory in four dimensions}",
    eprint = "1505.06029",
    archivePrefix = "arXiv",
    primaryClass = "gr-qc",
    reportNumber = "KEK-TH-1826, RUP-15-11",
    doi = "10.1007/JHEP07(2015)008",
    journal = "JHEP",
    volume = "07",
    pages = "008",
    year = "2015"
}

@article{Padilla:2012dx,
    author = "Padilla, Antonio and Sivanesan, Vishagan",
    title = "{Covariant multi-galileons and their generalisation}",
    eprint = "1210.4026",
    archivePrefix = "arXiv",
    primaryClass = "gr-qc",
    doi = "10.1007/JHEP04(2013)032",
    journal = "JHEP",
    volume = "04",
    pages = "032",
    year = "2013"
}

@article{Domenech:2023ryc,
    author = "Dom{\`e}nech, Guillem and Ganz, Alexander",
    title = "{Disformal symmetry in the Universe: mimetic gravity and beyond}",
    eprint = "2304.11035",
    archivePrefix = "arXiv",
    primaryClass = "gr-qc",
    doi = "10.1088/1475-7516/2023/08/046",
    journal = "JCAP",
    volume = "08",
    pages = "046",
    year = "2023"
}

@article{deRham:2016wji,
    author = "de Rham, Claudia and Matas, Andrew",
    title = "{Ostrogradsky in Theories with Multiple Fields}",
    eprint = "1604.08638",
    archivePrefix = "arXiv",
    primaryClass = "hep-th",
    doi = "10.1088/1475-7516/2016/06/041",
    journal = "JCAP",
    volume = "06",
    pages = "041",
    year = "2016"
}

@article{Crisostomi:2016czh,
    author = "Crisostomi, Marco and Koyama, Kazuya and Tasinato, Gianmassimo",
    title = "{Extended Scalar-Tensor Theories of Gravity}",
    eprint = "1602.03119",
    archivePrefix = "arXiv",
    primaryClass = "hep-th",
    doi = "10.1088/1475-7516/2016/04/044",
    journal = "JCAP",
    volume = "04",
    pages = "044",
    year = "2016"
}

@article{BenAchour:2016fzp,
    author = "Ben Achour, Jibril and Crisostomi, Marco and Koyama, Kazuya and Langlois, David and Noui, Karim and Tasinato, Gianmassimo",
    title = "{Degenerate higher order scalar-tensor theories beyond Horndeski up to cubic order}",
    eprint = "1608.08135",
    archivePrefix = "arXiv",
    primaryClass = "hep-th",
    doi = "10.1007/JHEP12(2016)100",
    journal = "JHEP",
    volume = "12",
    pages = "100",
    year = "2016"
}

@article{Dimopoulos:2005ac,
    author = "Dimopoulos, S. and Kachru, S. and McGreevy, J. and Wacker, Jay G.",
    title = "{N-flation}",
    eprint = "hep-th/0507205",
    archivePrefix = "arXiv",
    reportNumber = "SLAC-PUB-11016, SU-ITP-05-08",
    doi = "10.1088/1475-7516/2008/08/003",
    journal = "JCAP",
    volume = "08",
    pages = "003",
    year = "2008"
}

@article{Kanti:1999vt,
    author = "Kanti, Panagiota and Olive, Keith A.",
    title = "{On the realization of assisted inflation}",
    eprint = "hep-ph/9903524",
    archivePrefix = "arXiv",
    reportNumber = "UMN-TH-1756-99, TPI-MINN-99-18",
    doi = "10.1103/PhysRevD.60.043502",
    journal = "Phys. Rev. D",
    volume = "60",
    pages = "043502",
    year = "1999"
}

@article{Kanti:1999ie,
    author = "Kanti, Panagiota and Olive, Keith A.",
    title = "{Assisted chaotic inflation in higher dimensional theories}",
    eprint = "hep-ph/9906331",
    archivePrefix = "arXiv",
    reportNumber = "UMN-TH-1805-99, TPI-MINN-99-31",
    doi = "10.1016/S0370-2693(99)00982-X",
    journal = "Phys. Lett. B",
    volume = "464",
    pages = "192--198",
    year = "1999"
}

@article{Kobayashi:2013ina,
    author = "Kobayashi, Tsutomu and Tanahashi, Norihiro and Yamaguchi, Masahide",
    title = "{Multifield extension of $G$ inflation}",
    eprint = "1308.4798",
    archivePrefix = "arXiv",
    primaryClass = "hep-th",
    reportNumber = "RUP-13-7, IPMU13-0158",
    doi = "10.1103/PhysRevD.88.083504",
    journal = "Phys. Rev. D",
    volume = "88",
    number = "8",
    pages = "083504",
    year = "2013"
}

@article{Silverstein:2003hf,
    author = "Silverstein, Eva and Tong, David",
    title = "{Scalar speed limits and cosmology: Acceleration from D-cceleration}",
    eprint = "hep-th/0310221",
    archivePrefix = "arXiv",
    reportNumber = "SLAC-PUB-10211, MIT-CTP-3426, SU-ITP-3-29, NSF-KITP-03-87",
    doi = "10.1103/PhysRevD.70.103505",
    journal = "Phys. Rev. D",
    volume = "70",
    pages = "103505",
    year = "2004"
}

@article{Alishahiha:2004eh,
    author = "Alishahiha, Mohsen and Silverstein, Eva and Tong, David",
    title = "{DBI in the sky}",
    eprint = "hep-th/0404084",
    archivePrefix = "arXiv",
    reportNumber = "SLAC-PUB-10406, IPM-P-2004-016, MIT-CTP-3488, SU-ITP-4-15",
    doi = "10.1103/PhysRevD.70.123505",
    journal = "Phys. Rev. D",
    volume = "70",
    pages = "123505",
    year = "2004"
}

@article{Hinterbichler:2010xn,
    author = "Hinterbichler, Kurt and Trodden, Mark and Wesley, Daniel",
    title = "{Multi-field galileons and higher co-dimension branes}",
    eprint = "1008.1305",
    archivePrefix = "arXiv",
    primaryClass = "hep-th",
    doi = "10.1103/PhysRevD.82.124018",
    journal = "Phys. Rev. D",
    volume = "82",
    pages = "124018",
    year = "2010"
}

@article{Renaux-Petel:2011rmu,
    author = "Renaux-Petel, Sebastien and Mizuno, Shuntaro and Koyama, Kazuya",
    title = "{Primordial fluctuations and non-Gaussianities from multifield DBI Galileon inflation}",
    eprint = "1108.0305",
    archivePrefix = "arXiv",
    primaryClass = "astro-ph.CO",
    doi = "10.1088/1475-7516/2011/11/042",
    journal = "JCAP",
    volume = "11",
    pages = "042",
    year = "2011"
}

@article{deRham:2010eu,
    author = "de Rham, Claudia and Tolley, Andrew J.",
    title = "{DBI and the Galileon reunited}",
    eprint = "1003.5917",
    archivePrefix = "arXiv",
    primaryClass = "hep-th",
    doi = "10.1088/1475-7516/2010/05/015",
    journal = "JCAP",
    volume = "05",
    pages = "015",
    year = "2010"
}

@article{Ostrogradsky:1850fid,
    author = "Ostrogradsky, M.",
    title = "{M{\'e}moires sur les {\'e}quations diff{\'e}rentielles, relatives au probl{\`e}me des isop{\'e}rim{\`e}tres}",
    journal = "Mem. Acad. St. Petersbourg",
    volume = "6",
    number = "4",
    pages = "385--517",
    year = "1850"
}

@article{Horndeski:1974wa,
    author = "Horndeski, Gregory Walter",
    title = "{Second-order scalar-tensor field equations in a four-dimensional space}",
    doi = "10.1007/BF01807638",
    journal = "Int. J. Theor. Phys.",
    volume = "10",
    pages = "363--384",
    year = "1974"
}

@article{Deffayet:2011gz,
    author = "Deffayet, C. and Gao, Xian and Steer, D. A. and Zahariade, G.",
    title = "{From k-essence to generalised Galileons}",
    eprint = "1103.3260",
    archivePrefix = "arXiv",
    primaryClass = "hep-th",
    doi = "10.1103/PhysRevD.84.064039",
    journal = "Phys. Rev. D",
    volume = "84",
    pages = "064039",
    year = "2011"
}

@article{Gleyzes:2014dya,
    author = "Gleyzes, J\'er\^ome and Langlois, David and Piazza, Federico and Vernizzi, Filippo",
    title = "{Healthy theories beyond Horndeski}",
    eprint = "1404.6495",
    archivePrefix = "arXiv",
    primaryClass = "hep-th",
    doi = "10.1103/PhysRevLett.114.211101",
    journal = "Phys. Rev. Lett.",
    volume = "114",
    number = "21",
    pages = "211101",
    year = "2015"
}

@article{Gleyzes:2014qga,
    author = "Gleyzes, J\'er\^ome and Langlois, David and Piazza, Federico and Vernizzi, Filippo",
    title = "{Exploring gravitational theories beyond Horndeski}",
    eprint = "1408.1952",
    archivePrefix = "arXiv",
    primaryClass = "astro-ph.CO",
    doi = "10.1088/1475-7516/2015/02/018",
    journal = "JCAP",
    volume = "02",
    pages = "018",
    year = "2015"
}

@article{Langlois:2015cwa,
    author = "Langlois, David and Noui, Karim",
    title = "{Degenerate higher derivative theories beyond Horndeski: evading the Ostrogradski instability}",
    eprint = "1510.06930",
    archivePrefix = "arXiv",
    primaryClass = "gr-qc",
    doi = "10.1088/1475-7516/2016/02/034",
    journal = "JCAP",
    volume = "02",
    pages = "034",
    year = "2016"
}

@article{Langlois:2015skt,
    author = "Langlois, David and Noui, Karim",
    title = "{Hamiltonian analysis of higher derivative scalar-tensor theories}",
    eprint = "1512.06820",
    archivePrefix = "arXiv",
    primaryClass = "gr-qc",
    doi = "10.1088/1475-7516/2016/07/016",
    journal = "JCAP",
    volume = "07",
    pages = "016",
    year = "2016"
}

@article{Babichev:2021bim,
    author = "Babichev, Eugeny and Izumi, Keisuke and Tanahashi, Norihiro and Yamaguchi, Masahide",
    title = "{Invertibility conditions for field transformations with derivatives: Toward extensions of disformal transformation with higher derivatives}",
    eprint = "2109.00912",
    archivePrefix = "arXiv",
    primaryClass = "hep-th",
    doi = "10.1093/ptep/ptab151",
    journal = "PTEP",
    volume = "2022",
    number = "1",
    pages = "013A01",
    year = "2022"
}

@article{Babichev:2024eoh,
    author = "Babichev, Eugeny and Izumi, Keisuke and Noui, Karim and Tanahashi, Norihiro and Yamaguchi, Masahide",
    title = "{Generalization of conformal-disformal transformations of the metric in scalar-tensor theories}",
    eprint = "2405.13126",
    archivePrefix = "arXiv",
    primaryClass = "gr-qc",
    reportNumber = "KUNS-3003",
    doi = "10.1103/PhysRevD.110.064063",
    journal = "Phys. Rev. D",
    volume = "110",
    number = "6",
    pages = "064063",
    year = "2024"
}

@article{Domenech:2019syf,
    author = "Dom{\`e}nech, Guillem and Naruko, Atsushi and Sasaki, Misao and Wetterich, Christof",
    title = "{Could the black hole singularity be a field singularity?}",
    eprint = "1912.02845",
    archivePrefix = "arXiv",
    primaryClass = "gr-qc",
    reportNumber = "TU1084, YITP-19-53",
    doi = "10.1142/S0218271820500261",
    journal = "Int. J. Mod. Phys. D",
    volume = "29",
    number = "03",
    pages = "2050026",
    year = "2020"
}

@article{Naruko:2022vuh,
    author = "Naruko, Atsushi and Saito, Ryo and Tanahashi, Norihiro and Yamauchi, Daisuke",
    title = "{Ostrogradsky mode in scalar{\textendash}tensor theories with higher-order derivative couplings to matter}",
    eprint = "2209.02252",
    archivePrefix = "arXiv",
    primaryClass = "gr-qc",
    reportNumber = "YITP-22-94",
    doi = "10.1093/ptep/ptad049",
    journal = "PTEP",
    volume = "2023",
    number = "5",
    pages = "053E02",
    year = "2023"
}

@article{DeFelice:2018ewo,
    author = "De Felice, Antonio and Langlois, David and Mukohyama, Shinji and Noui, Karim and Wang, Anzhong",
    title = "{Generalized instantaneous modes in higher-order scalar-tensor theories}",
    eprint = "1803.06241",
    archivePrefix = "arXiv",
    primaryClass = "hep-th",
    doi = "10.1103/PhysRevD.98.084024",
    journal = "Phys. Rev. D",
    volume = "98",
    number = "8",
    pages = "084024",
    year = "2018"
}

@article{DeFelice:2021hps,
    author = "De Felice, Antonio and Mukohyama, Shinji and Takahashi, Kazufumi",
    title = "{Nonlinear definition of the shadowy mode in higher-order scalar-tensor theories}",
    eprint = "2110.03194",
    archivePrefix = "arXiv",
    primaryClass = "gr-qc",
    reportNumber = "YITP-21-104, IPMU21-0061",
    doi = "10.1088/1475-7516/2021/12/020",
    journal = "JCAP",
    volume = "12",
    number = "12",
    pages = "020",
    year = "2021"
}

@article{auth:Domenech_consistency_conditions,
    author = "Dom{\`e}nech, Guillem and Ganz, Alexander and Tsampodimos, Apostolos",
    title = "{On the consistent disformal couplings to fermions}",
    eprint = "2510.07419",
    archivePrefix = "arXiv",
    primaryClass = "hep-th",
    doi = "10.1088/1475-7516/2026/02/045",
    journal = "JCAP",
    volume = "02",
    pages = "045",
    year = "2026"
}

@article{BouzariNezhad:2026zsj,
    author = "Bouzari Nezhad, Hamed",
    title = "{Removing Ostrogradsky modes in multi-field higher-order scalar-tensor theories}",
    eprint = "2606.23734",
    archivePrefix = "arXiv",
    primaryClass = "gr-qc",
    month = "6",
    year = "2026"
}

@article{Katayama:2025hnd,
    author = "Katayama, Tomoki",
    title = "{A New Definition of Horndeski Theory and the Possibility of Multiple Scalar Field Extensions}",
    eprint = "2511.15423",
    archivePrefix = "arXiv",
    primaryClass = "gr-qc",
    reportNumber = "KEK-Cosmo-0398, KEK-TH-2777",
    month = "11",
    year = "2025"
}

@article{auth:Deffayet_GarciaSaenz,
    author = "Deffayet, C{\'e}dric and Garcia-Saenz, Sebastian",
    title = "{Degeneracy, matter coupling, and disformal transformations in scalar-tensor theories}",
    eprint = "2004.11619",
    archivePrefix = "arXiv",
    primaryClass = "hep-th",
    doi = "10.1103/PhysRevD.102.064037",
    journal = "Phys. Rev. D",
    volume = "102",
    number = "6",
    pages = "064037",
    year = "2020"
}

@article{auth:Gao_Yao_SCG,
    author = "Gao, Xian and Yao, Zhi-Bang",
    title = "{Spatially covariant gravity with velocity of the lapse function: the Hamiltonian analysis}",
    eprint = "1806.02811",
    archivePrefix = "arXiv",
    primaryClass = "gr-qc",
    doi = "10.1088/1475-7516/2019/05/024",
    journal = "JCAP",
    volume = "05",
    pages = "024",
    year = "2019"
}

@book{auth:DeWitt_Supermanifolds,
    author = "DeWitt, Bryce S.",
    title = "{Supermanifolds}",
    series = "Cambridge Monographs on Mathematical Physics",
    edition = "2",
    publisher = "Cambridge University Press",
    address = "Cambridge",
    year = "1992"
}

@book{auth:Henneaux_Teitelboim,
    author = "Henneaux, Marc and Teitelboim, Claudio",
    title = "{Quantization of gauge systems}",
    publisher = "Princeton University Press",
    address = "Princeton",
    year = "1992"
}

@article{auth:Kimura_Boson_Fermion,
    author = "Kimura, Rampei and Sakakihara, Yuki and Yamaguchi, Masahide",
    title = "{Ghost free systems with coexisting bosons and fermions}",
    eprint = "1704.02717",
    archivePrefix = "arXiv",
    primaryClass = "hep-th",
    doi = "10.1103/PhysRevD.96.044015",
    journal = "Phys. Rev. D",
    volume = "96",
    number = "4",
    pages = "044015",
    year = "2017"
}

@article{auth:Kimura_Scalar_Fermion,
    author = "Kimura, Rampei and Sakakihara, Yuki and Yamaguchi, Masahide",
    title = "{Ghost-free scalar-fermion interactions}",
    eprint = "1805.10963",
    archivePrefix = "arXiv",
    primaryClass = "hep-th",
    doi = "10.1103/PhysRevD.98.044043",
    journal = "Phys. Rev. D",
    volume = "98",
    number = "4",
    pages = "044043",
    year = "2018"
}

\end{document}